\documentclass[preprint,3p,12pt]{elsarticle}
\usepackage{mathrsfs}
\usepackage{amsmath}
\usepackage{stmaryrd}
\usepackage{bbding}
\usepackage{dcolumn}
\usepackage{amsfonts}
\usepackage{amssymb}
\usepackage{psfrag}
\usepackage{wrapfig}
\usepackage{makeidx}
\usepackage{bm}
\usepackage{epsf}
\usepackage{epsfig}
\usepackage{setspace}
\usepackage{epstopdf}
\usepackage{psfrag}
\usepackage{color}
\usepackage{scalerel}  
\usepackage{hyperref}
\usepackage{subfigure}
\hypersetup{
	colorlinks = true,
	urlcolor   = blue,
	citecolor  = black,
}
\usepackage{lineno}

\usepackage{graphicx}
\usepackage{caption}
\usepackage{overpic}

\newcommand\reallywidehat[1]{\arraycolsep=0pt\relax
	\begin{array}{c}
		\stretchto{
			\scaleto{
				\scalerel*[\widthof{\ensuremath{#1}}]{\kern-.5pt\bigwedge\kern-.5pt}
				{\rule[-\textheight/2]{1ex}{\textheight}} 
			}{\textheight} %
		}{0.5ex}\\           
		#1\\                 
		\rule{-1ex}{0ex}
	\end{array}
}
\begin{document}
\title{High-order Gas-kinetic Schemes with Non-compact and Compact Reconstruction
for Implicit Large Eddy Simulation}
\author[HKUST1,SUSTech1]{Wenjin Zhao}

\author[SUSTech2]{Guiyu Cao}

\author[SUSTech1]{Jianchun Wang \corref{cor}}
\ead{wangjc@sustech.edu.cn}
\author[HKUST1,HKUST2,HKUST3]{Kun Xu}

\address[HKUST1]{Department of Mechanical and Aerospace Engineering, Hong Kong University of Science and Technology, Clear Water Bay, Kowloon, HongKong}
\address[SUSTech1]{Department of Mechanics and Aerospace Engineering, Southern University of Science and Technology, Shenzhen 518055, P.R.China}
\address[SUSTech2]{Academy for Advanced Interdisciplinary Studies, Southern University of Science and Technology, Shenzhen, P.R.China}
\address[HKUST2]{Department of Mathematics, Hong Kong University of Science and Technology, Clear Water Bay, Kowloon, Hong Kong}
\address[HKUST3]{Shenzhen Research Institute, Hong Kong University of Science and Technology, Shenzhen, Guangdong 518057, PR China}
\cortext[cor]{Corresponding author}

\begin{abstract}
High-order gas-kinetic scheme (HGKS) with $5$th-order non-compact reconstruction has been well implemented for implicit large eddy simulation (ILES) in nearly incompressible turbulent channel flows.
In this study, the HGKS with higher-order non-compact reconstruction and compact reconstruction will be validated in turbulence simulation. For higher-order non-compact reconstruction, $7$th-order normal reconstruction and tangential reconstruction are implemented. In terms of compact reconstruction, $5$th-order normal reconstruction is adopted. 
Current work aims to show the benefits of high-order non-compact reconstruction and compact reconstruction for ILES. 
The accuracy of HGKS is verified by numerical simulation of three-dimensional advection of density perturbation. For the non-compact 7th-order scheme, 16 Gaussian points are required on the cell-interface to preserve the order of accuracy.
Then, HGKS with non-compact and compact reconstruction is used in the three-dimensional Taylor-Green vortex (TGV) problem and  turbulent channel flows.
Accurate ILES solutions have been obtained from HGKS. In terms of the physical modeling underlying the numerical algorithms, the compact reconstruction has the consistent physical and numerical domains of dependence without employing additional information from cells which have no any direct physical connection with the targeted cell. The compact GKS shows a favorable performance for turbulence simulation in resolving the multi-scale structures.
\end{abstract}


\maketitle

\section{\label{sec:level1} Introuction}

Turbulence is a universal physical phenomenon and the research of turbulence is of great significance to industry \cite{pope2001turbulent}. Due to its multi-scale nature in space and time, it is a challenge to balance the accuracy requirements and computational costs in turbulence simulations.
Currently, there are four main approaches in numerical simulation of turbulent flow: direct numerical simulation (DNS) \cite{kim1987turbulence,moin1998direct,moser1999direct,wang2020effect}, Reynolds averaged Navier-Stokes (RANS) simulation  \cite{chou1945velocity, jones1972prediction, spalart1992one, menter1994two}, large eddy simulation (LES) \cite{smagorinsky1963general, ferziger1979evaluation, germano1991dynamic,vreman1996large, nicoud1999subgrid, vreman2004eddy, sagaut2006large, pantano2008approach, rozema2015minimum}, and hybrid RANS/LES methods(HRLM) \cite{spalart2009detached}.
Implicit LES (ILES) was proposed in 1992 \cite{boris1992new}, in which the numerical dissipation can be used to replace the SGS dissipation in the turbulence simulation. The ILES approach can overcome the problem of over-dissipation that arises in explicit LES models \cite{bazilevs2007variational,kokkinakis2015implicit}.
 Christer Fureby and Fernando F. Grinstein constructed the monotonically integrated LES (MILES) \cite{2002Large}. In this method, intrinsic nonlinear high-frequency filters built into the convection discretization provide the implicit SGS models, which are coupled naturally to the resolvable scales of the flow.
Y. Bazilevs $et\,al.$ developed variational multiscale residual-based turbulence modeling \cite{bazilevs2007variational}, which can be derived completely from the incompressible Navier–Stokes equations without employing any ad hoc devices, such as eddy viscosities in the traditional LES models.
  I.W. Kokkinakis and D. Drikakis used several high-resolution and high-order finite volume schemes for the simulation of weakly compressible  turbulent channel flow and verified the advantage of the higher-order scheme in ILES \cite{kokkinakis2015implicit}.

%
The gas-kinetic scheme (GKS) \cite{xu2001gas,xu2015direct} is a finite volume method based on the Bhatnagar--Gross--Krook (BGK) model \cite{bhatnagar1954model} for the construction of gas evolution model at a cell interface.
In recent years, in conjunction with weighted essentially non-oscillatory (WENO) reconstruction and two-stage fourth-order (S2O4) temporal discretization \cite{li2016two,pan2016efficient}, GKS has achieved great success in flow simulations with high temporal and spatial resolutions. High-order GKS (HGKS) coupled with the turbulence model has been effectively applied in RANS and LES of turbulence \cite{cao2019implicit, pan2021fourth}. 
The performance of HGKS with parallel computation has been investigated in the
DNS of turbulence flows \cite{cao2019three,cao2021highdns}. 
The comparison of the performance for ILES and LES with 5th-order GKS has been studied, which shows that the high-order GKS can provide appropriate numerical dissipation and is suitable for ILES of turbulence \cite{zhao2021high}. 
Additionally, high-order gas-kinetic scheme in general curvilinear coordinate has been developed for ILES of compressible wall-bounded turbulent flows \cite{cao2021high}.
The compact GKS (CGKS) has been developed in recent years and successfully applied to compressible flow simulation, which exhibits superiority compared with the non-compact scheme \cite{ji2018compact,zhao2019compact,zhao2020acoustic,ji2021gradient}.
The GKS provides a time-accurate solution at a cell interface based on the high-order gas evolution model. From the time-accurate solution, both time-accurate flux function and flow variables can be obtained for updating cell averaged conservative flow variables and gradients. Based on cell-averaged flow variables and gradients, the compact reconstruction in GKS can be obtained.
Compared with the non-compact scheme, the compact scheme can achieve higher-order accuracy with the same stencils. Additionally, the stencils used in compact scheme is less than that used in the non-compact scheme for the same order of accuracy.

In this study, we develop the GKS with 7th-order non-compact reconstruction  in the normal and tangential directions, then investigate the effect of improving the reconstruction order on turbulence simulation. Following the idea of the compact GKS \cite{ji2018compact}, we also develop the HGKS with 5th-order compact reconstruction in the normal direction and investigate the effect of compact reconstruction of GKS on turbulence simulation. The three-dimensional advection of density perturbation is tested to verify the accuracy of different schemes. The three-dimensional Taylor-Green vortex and turbulent channel flows are simulated by HGKS. For the high-order scheme, the numerical dissipation is reduced, and the resolution is increased. The compact reconstruction has the consistent physical and numerical domains of dependence \cite{ji2018compact,zhao2019compact}. 
The compact GKS has a favorable performance for turbulence simulation in resolving the multi-scale structure.



The remainder of this paper is organized as follows: Section \ref{sec2} introduces the HGKS. Section \ref{sec3} presents the HGKS with the higher-order non-compact and compact reconstruction. In Section \ref{sec4}, the accuracy tests for the schemes are conducted. In Section \ref{sec5}, the schemes are applied in TGV and  turbulent channel flows. Section \ref{sec6} is the discussion and conclusion.

\section{\label{sec2}High-order gas kinetic scheme}

The three-dimensional BGK equation can be written as \cite{bhatnagar1954model}
\begin{equation}\label{boltzman_bgk}
    \frac{\partial f}{\partial t} + u\frac{\partial{f}}{\partial x}+ v\frac{\partial{f}}{\partial y}+ w\frac{\partial{f}}{\partial z} = \frac{g-f}{\tau},
\end{equation}
where $u,v,w$ is the particle velocity in three dimensions;
$g$ denotes the equilibrium state in the Maxwellian distribution and is approached by $f$; and collision time $\tau$ denotes the average time interval between successive particle collisions for the same particle.
The collision term satisfies the compatibility condition
\begin{equation*} \label{bgk_compatibility}
\int \frac{g-f}{\tau} \bm{\psi} \text{d}\Xi=0,
\end{equation*}
where $\bm{\psi} = (1, u, v, w, \frac{1}{2}(u^2 + v^2 + w^2 + \xi^2))^T$, $\xi^2=\xi_1^2+...+\xi_N^2$, and $\text{d}\Xi=\text{d}u\text{d}v\text{d}w\text{d}\xi_1…\text{d}\xi_{N}$. Here, N is the internal degree of freedom and relates to the specific heat $\gamma$, with $N=(5-3\gamma)(\gamma-1)$. 

The time-dependent gas distribution function at the cell interface can be expressed as the integral solution of the BGK equation as follows \cite{xu2015direct}:

\begin{equation} \label{gks_formalsolution}
\begin{aligned}
f(\boldsymbol{x}_{i+1/2,j_m,k_n},t,\boldsymbol{u},\xi)&=\frac{1}{\tau}\int_0^t
g(\boldsymbol{x}',t',\boldsymbol{u},\xi)e^{-(t-t')/\tau}\text{d} t'\\&+e^{-t/\tau}f_0(-\boldsymbol{u}t,\xi),
\end{aligned}
\end{equation}
where $(y_{j_m},z_{k_n})$ is the Gaussian quadrature point of the cell interface $\overline{y}_j \times \overline{z}_k$; $\boldsymbol{u}=(u,v,w)^T$ is the particle velocity in three dimensions; $\boldsymbol{x}'=\boldsymbol{x}_{i+1/2,j_m,k_n}-\boldsymbol{u}(t-t')$ is the particle trajectory; $f_0$ is the initial gas distribution function at the beginning of each time step; and $g$ is the corresponding equilibrium state.

Here, $f_0$ is assumed to be
\begin{equation}\label{gks_f0}
\begin{aligned}
f_0 =
\begin{cases}
g_l [1 +  (a_l x + b_l y + c_l z) - \tau (a_l u + b_l v + c_l w + A_l)],x\leq 0, \\
g_r [1 +  (a_r x + b_r y + c_r z) - \tau (a_r u + b_r v + c_r w + A_r)],x > 0,
\end{cases}
\end{aligned}
\end{equation}
where $g_l$ and $g_r$ are the Maxwellian distributions at the two sides of the cell interface, $a_{l,r},b_{l,r},c_{l,r}$ correspond to the coefficients in spatial derivatives in the expansion of a Maxwellian and $A_{l,r}$ correspond to the coefficients in temporal derivatives.

Further, $g$ can be expressed as
\begin{equation}\label{gks_g}
\begin{aligned}
g = g_c(1 + \overline{a} x + \overline{b} y + \overline{c} z + \overline{A} t),
\end{aligned}
\end{equation}
where $g_c$ is the initial equilibrium state located at the cell interface, which can be determined from the compatibility condition
\begin{align}\label{g0_compatibility}
 \int  g_c\bm{\psi} \text{d} \Xi = \int_{u>0}  g_l\bm{\psi} \text{d}  \Xi + \int_{u<0}  g_r\bm{\psi} \text{d} \Xi.
\end{align}
Here, $g_l$ and $g_r$ are the initial equilibrium gas distribution functions on both sides of the cell interface. The spatial microscopic coefficients, $i.e.$, $a_{k}, b_{k}, c_{k}, \overline{a}, \overline{b}$ and $\overline{c}$ can be calculated from the slope of macro conserved quantities at the two sides of the cell interface $\bm{Q}^k$, and the corresponding equilibrium state $\bm{Q}^c$
\begin{align*}
\displaystyle \langle a_{k}\rangle=\frac{\partial
	\bm{Q}^{k}}{\partial x}, \langle
b_{k}\rangle=\frac{\partial \bm{Q}^{k}}{\partial y},
\langle c_{k}\rangle&=\frac{\partial \bm{Q}^{k}}{\partial
	z}
 \\ \displaystyle
\langle\overline{a}\rangle=\frac{\partial \bm{Q}^{c}}{\partial
	x}, \langle\overline{b}\rangle=\frac{\partial
	\bm{Q}^{c}}{\partial y},
\langle\overline{c} \rangle&=\frac{\partial \bm{Q}^{c}}{\partial
	z},
\end{align*}
where $k=l,r$.
 The temporal microscopic coefficients, $i.e.$, $A_{k}$ and $\overline{A}$ can be determined from the compatibility condition \cite{pan2016efficient}
\begin{align*}
\langle a_{k}u + b_{k} v + c_{k}w + A^{k}\rangle=0,
\langle\overline{a}u + \overline{b} v + \overline{c} w + \overline{A}\rangle = 0,
\end{align*}
 where $k=l,r$ and $\langle...\rangle$ are the moments of the
equilibrium $g$ and defined by $\rho \langle...\rangle=\int g (...) \bm{\psi} \text{d}\Xi$.

As illustrated above, the equilibrium distribution functions $g_l$, $g_r$ and $g_c$, as well as the slopes $a_{k}, b_{k}, c_{k},$ $\overline{a}, \overline{b}$, and $\overline{c}$ can be obtained from the macroscopic conservative flow variables and their slopes around a cell interface. The compact and non-compact high-order reconstruction strategies are used to get the macroscopic variables and their slopes at the Gaussian points around the cell interface. 	Substituting Eqs. (\ref{gks_f0}) and (\ref{gks_g}) into the formal solution of Eq. (\ref{gks_formalsolution}), the second-order gas distribution function at the cell interface can be expressed as \cite{xu2015direct}

\begin{align}\label{formalsolution_neq}
f(\boldsymbol{x}_{i+1/2,j_m,k_n},t,\boldsymbol{u},\xi) &= (1-e^{-t/\tau})g_c\nonumber\\
&+((t+\tau)e^{-t/\tau}-\tau)(\overline{a} u + \overline{b} v + \overline{c} w)g_c\nonumber\\
&+(t-\tau+\tau e^{-t/\tau}){\bar{A}} g_c \nonumber\\
&+e^{-t/\tau}g_l[1-(\tau+t)(a_l u + b_l v + c_l w)-\tau
A^l)]H(u) \nonumber\\
&+e^{-t/\tau}g_r[1-(\tau+t)(a_r u + b_r v + c_r w)-\tau A^r)](1-H(u)).
\end{align}

In the smooth-flow region, for the continuous flow variables at the cell interface,  the gas distribution function can be simplified as
\begin{equation}\label{formalsolution_smooth}
\begin{aligned}
f(\boldsymbol{x}_{i+1/2,j_m,k_n},t,\boldsymbol{u},\xi) =  g_c [1 - \tau(\overline{a} u + \overline{b} v + \overline{c} w + \overline{A}) + t\overline{A}].
\end{aligned}
\end{equation}
Taking the moments of the BGK equation ($i.e.$, Eq. (1)) and integrating with respect to space, the
finite volume scheme can be expressed as
\begin{align}\label{hgks_semi}
    \frac{\text{d}(\bm{Q}_{ijk})}{\text{d}t}=\mathcal{L}(\bm{Q}_{ijk}).
\end{align}
Here, the $\mathcal{L}$ operator is defined as
\begin{equation}\label{hgks_L}
\begin{aligned}
    \mathcal{L}(\bm{Q}_{ijk})= - \frac{1}{|\Omega_{ijk}|} [&\int_{\overline{y_j} \times \overline{z_k}} (\bm{F}_{i + 1/2, j, k} - \bm{F}_{i - 1/2, j, k}) \text{d}y \text{d}z \\
    + &\int_{\overline{x_i} \times \overline{z_k}} (\bm{G}_{i, j + 1/2, k} - \bm{G}_{i, j - 1/2, k}) \text{d}x \text{d}z \\
    + &\int_{\overline{x_i} \times \overline{y_j}} (\bm{H}_{i, j, k + 1/2} - \bm{H}_{i, j, k - 1/2}) \text{d}x \text{d}y],
   \end{aligned}
\end{equation}
where $|\Omega_{ijk}|$ is the control volume with $\overline{x}_i=[x_i-\Delta x/2,x_i+\Delta x/2], \overline{y}_j=[y_j-\Delta y/2,y_j+\Delta y/2]$, and $\overline{z}_k=[z_k-\Delta z/2,z_k+\Delta z/2]$, and $\bm{F},\bm{G}$, and $\bm{H}$ denote the fluxes of the conservative flow variables in three dimensions.

Taking the time-dependent numerical flux in the $x$-direction as an example \cite{cao2021high},
\begin{equation}\label{flux_integration}
\begin{aligned}
\int_{\overline{y_j} \times \overline{z_k}} \bm{F}_{i + 1/2, j, k} \text{d}y \text{d}z = \sum_{m,n=1}^c\omega_{mn}
\int \bm{\psi} u
f(\bm{x}_{i+1/2,j_m,k_n},t,\bm{u},\xi)\text{d}\Xi\Delta y\Delta z,
   \end{aligned}
\end{equation}
where $\bm{F}=(F_{\rho},F_{\rho U},F_{\rho V},F_{\rho W},F_{\rho  E})$ denote the fluxes of the conservative flow variables, $\omega_{mn}$ is the quadrature weight, $(y_{j_m},z_{k_n})$ is the Gaussian quadrature point of the cell interface $\overline{y}_j \times \overline{z}_k$, and $c=2$ and $4$ correspond to 4 and 16 Gaussian points used in the cell-interface, respectively.
The positions and weights for the Gaussian points $x_i$ in one-dimension are shown in Table \ref{gaussian}, where the width of the cell is $\Delta x$, and the coordinate of the cell center-point is 0. It can be easily extended to two-dimension through those parameters.
 For temporal updating, S2O4 time-accurate discretization is adopted. Further details on time-accurate discretization can be found in previous work \cite{pan2016efficient}.

 \begin{table}[!h]
	\caption{\label{gaussian} The positions and weights for the Gaussian points distribution}
	\vspace{3mm}
	\centering
\begin{tabular}{ccc}
\hline
\hline
Gaussian points & Coordinates & Weights \\
  \hline
  $x_1$ & $\frac{-1}{2\sqrt{3}} \Delta x$ &$\frac {1}{2} $\\

 $x_2$  &  $\frac{1}{2\sqrt{3}} \Delta x$ &$\frac{1}{2}$ \\
\hline
  $x_1$  & $-\sqrt{\frac{15+2 \sqrt{30}}{140}} \Delta x   $ & $\frac{18-\sqrt{30}}{72}$ \\

 $x_2$  &  $-\sqrt{\frac{15-2 \sqrt{30}}{140}} \Delta x   $ & $\frac{18+\sqrt{30}}{72}$ \\
  $x_3$  & $\quad \sqrt{\frac{15-2 \sqrt{30}}{140}} \Delta x   $ & $\frac{18+\sqrt{30}}{72}$\\

 $x_4$  &  $ \quad \sqrt{\frac{15+2 \sqrt{30}}{140}} \Delta x   $ & $\frac{18-\sqrt{30}}{72}$ \\

\hline
\hline
\end{tabular}
\vspace{-2mm}
\end{table}

\section{\label{sec3} Three-dimensional non-compact high-order reconstruction and compact reconstruction}

\subsection{\label{sec:nonc_recon} High-order non-compact reconstruction in three-dimension}
The reconstruction order of HGKS for turbulence simulation in previous work is mainly 5th-order \cite{cao2021highdns,zhao2021high}. In the following, we introduce the reconstruction method that improves the order to 7th-order.
To get the values of macroscopic conserved quantities and their slopes which are used for the construction of smooth flux at the Gaussian points in the cell interfaces, the direction by direction reconstruction strategy is applied on rectangular meshes \cite{ji2019high}. The details are illustrated as follows:

Step 1:
In the normal direction, the left and right face averaged values at the cell interface can be reconstructed by 7th-order WENO reconstruction \cite{2016An} with the seven cell averaged values as the sub-stencils. For the smooth flow cases, the linear weights in WENO are used. The face averaged values in the equilibrium state can be obtained by the compatibility condition.
After that, a linear sixth-order polynomial can be constructed using the cell averaged values, and the first-order derivative for face averaged values can be calculated from the polynomial, as shown in Section \ref{7face}.

Step 2:
	In the horizontal direction, a linear sixth-order polynomial can be constructed  using the face averaged values
	$(Q^c)_{j-l ,k}, l=-3,...,3$,
	and the line averaged values and derivatives
	$(\overline{Q}^c)_{j \pm 1/2,k}$, $(\overline{Q}^c_y)_{j \pm1/2,k}$
	at $y=y_{j_m}$ can be obtained from the polynomial, as shown in Section \ref{7cell}.
	
	Similarly we construct a linear sixth-order polynomial using the face averaged derivatives
	$(Q^c_x)_{j-l ,k}, l=-3,...,3$. 
	Then the derivatives for line averaged values $(\overline{Q}^c_x)_{j \pm1/2,k}$ at $y=y_{j_m}$ can be obtained.
	
Step 3:
	In vertical direction, similarly, a linear sixth-order polynomial can be constructed by using the line averaged values $(\overline{Q}^c)_{j_m,k+l},l=-3,...,3$.
 Then the point values and derivatives $(\dot{Q}^c)_{j_m,k_n},(\dot{Q}^c_z)_{j_m,k_n} $
	at the Gaussian points can be obtained. 
	The spatial derivatives at the Gaussian point $	(\dot{Q}^c_x)_{j_m,k_n},(\dot{Q}^c_y)_{j_m,k_n}$
	can be obtained in the same way.

\subsubsection{Algorithm for linear 7th-order reconstruction at cell interface}\label{7face}
To fully  determine the slopes of the equilibrium state across the cell interface, the variables across the cell interface $Q^{c}(x)$ are expanded as \cite{ji2019high}
\begin{equation}
\begin{aligned}
Q^{c}(x)&=Q_{i+1/2}^{c}+S_1(x-x_{i+1/2})+\frac{1}{2}S_2(x-x_{i+1/2})^2+\frac{1}{6}S_3(x-x_{i+1/2})^3\\&+\frac{1}{24}S_4(x-x_{i+1/2})^4+\frac{1}{120}S_5(x-x_{i+1/2})^5+\frac{1}{720}S_6(x-x_{i+1/2})^6,
\end{aligned}
\end{equation}
where $Q_{i+1/2}^{c}$ are the variables in equilibrium state at cell interface $x=x_{i+1/2}$. 
With the following conditions,
\begin{equation}
\begin{aligned}
\int_{I_{i+k}} Q^{c}(x) \text{d} x=Q_{i+k}, k=-2,...,3,
\end{aligned}
\end{equation}
the derivatives are determined  by $(Q_x^{c})_{i+1/2}=S_1.$

\subsubsection{Algorithm for linear 7th-order reconstruction at cell center}\label{7cell}
For the reconstruction of equilibrium state in horizontal and vertical directions we use the smooth reconstruction. The algorithm for smooth 7th-order reconstruction is explicated as follows. 
Firstly, we construct a 7th-order polynomial expansion at the cell center by using 7 sub-stencils 
\begin{equation}
\begin{aligned}
Q^{c}(x)&=Q_{i}^{c}+S_1(x-x_{i})+\frac{1}{2}S_2(x-x_{i})^2+\frac{1}{6}S_3(x-x_{i})^3\\ &+\frac{1}{24}S_4(x-x_{i})^4+\frac{1}{120}S_5(x-x_{i})^5+\frac{1}{720}S_6(x-x_{i})^6,
\end{aligned}
\end{equation}
where $Q^{c}(x)$ are the variables in equilibrium state at coordinate $x$, and $Q_{i}^{c}$ are the variables in equilibrium state at cell center $x=x_i$.

With the following conditions,
\begin{equation}
\begin{aligned}
\int_{I_{i+k}} Q^{c}(x) \text{d} x=Q_{i+k}, k=-3,...,3,
\end{aligned}
\end{equation}
the coefficients of polynomials can be obtained.
From the polynomial we can calculate the point values and spatial derivative values at the Gauss points.

\begin{figure*}

        \includegraphics[height=1.3\textwidth]{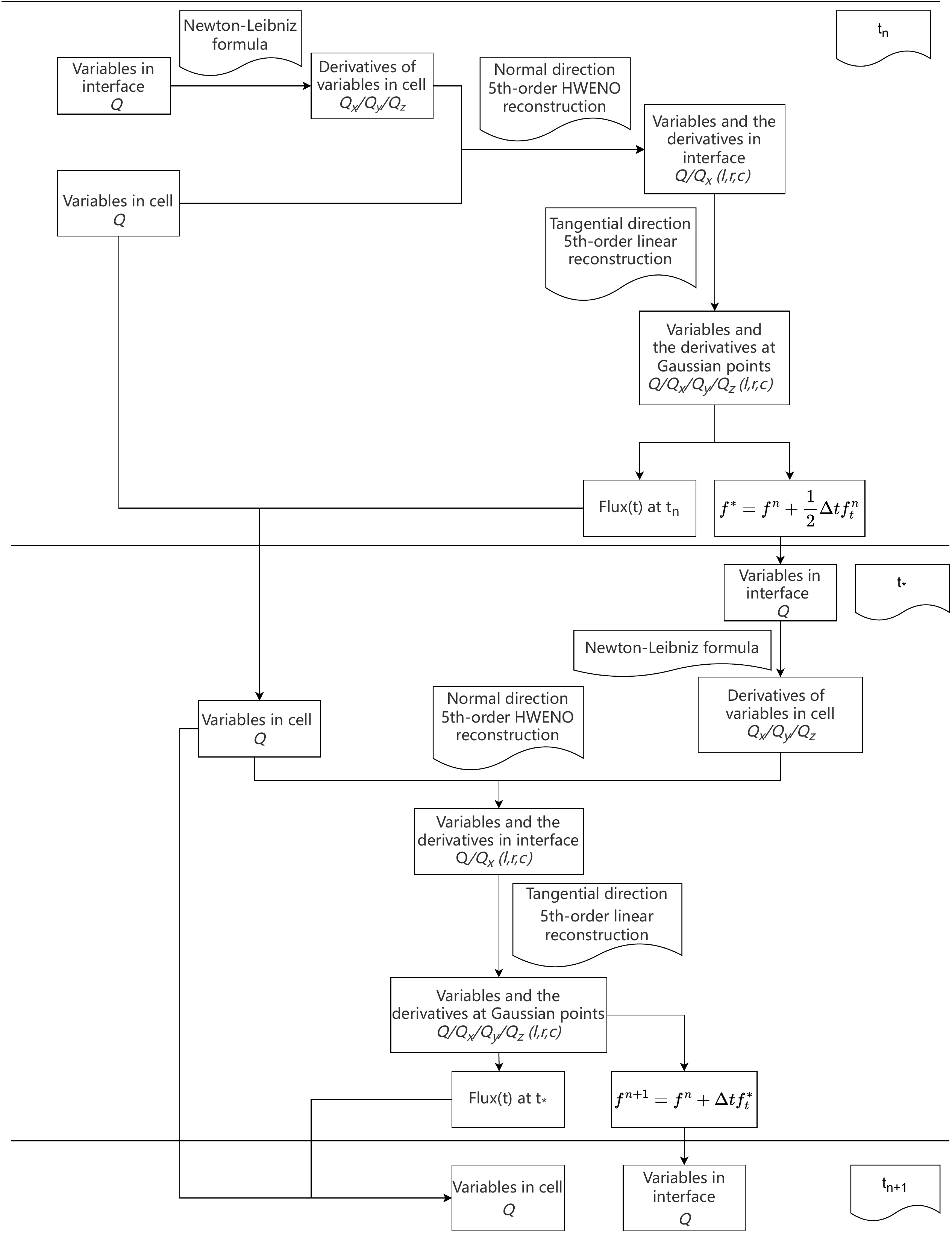}
  
\caption{\label{cgks} Algorithm for compact GKS.}
   
\end{figure*}
\begin{figure*}
   \centering
        \includegraphics[height=0.35\textwidth]{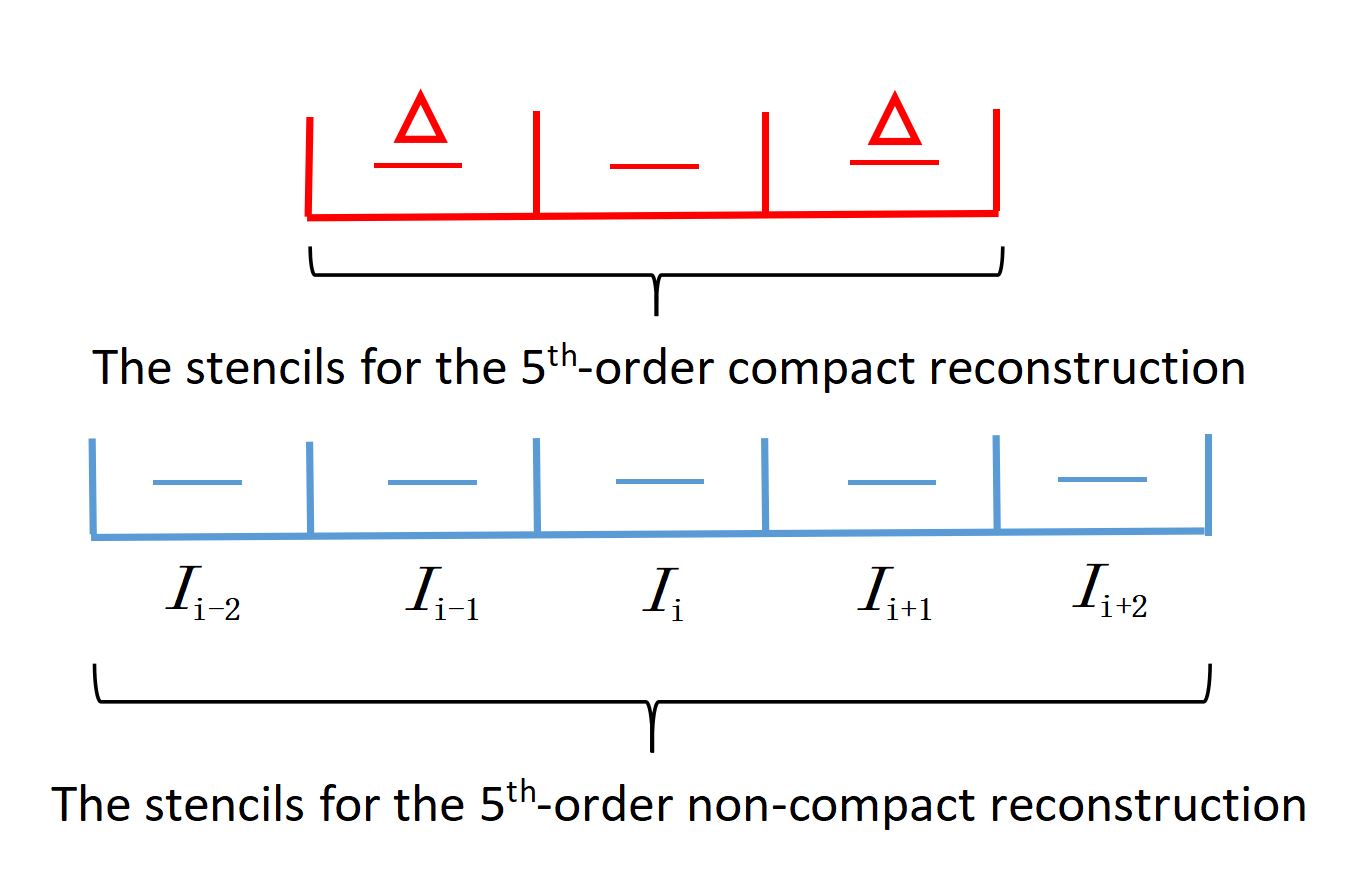}

\caption{\label{moban} The stencils for the non-compact and compact reconstruction}

\end{figure*}

\subsection{\label{sec:cgks} High-order compact reconstruction in three-dimension}
Following the idea of the compact GKS \cite{ji2018compact}, the algorithm for HGKS with 5th-order compact reconstruction in the normal direction can be illustrated in Fig. \ref{cgks}, and described as follows:

Step1: By applying the Newton-Leibniz formula as shown in Eq. (\ref{nlf}) for the conservative variables in the cell-interface, we can obtain the derivatives for conservative variables in the cell. 
In the normal direction, based on the cell averages and cell averaged derivatives, through applying the 5th-order Hermite weighted essentially non-oscillatory (HWENO) reconstruction \cite{2004Hermite}, the face averages $Q^l$ and $Q^r$ over cell interface can be obtained. For the smooth flow cases, the linear weights in HWENO are used. Then by the compatibility conditions, we get the equilibrium face average $Q^c$. The equilibrium face-averaged derivatives $Q_x$ are obtained by the linear 4th-order polynomial.
In tangential directions, we still use the 5th-order linear reconstruction, variables and the derivatives at the Gaussian points can be obtained. 
Then we can calculate the time dependent flux at $t_n$ at cell interface. Additionally, the temporal evolution for the interface values at time step $t_*$ can also be obtained. For the second-order GKS flux used in this research, the temporal evolution for the interface values \cite{ji2018compact} can be expressed as $f^*=f^n+ \frac {1}{2} \Delta t f^n_t$.

\begin{equation} \label{nlf}
\left(Q_{x}\right)_{i}=\frac{1}{\Delta x} \int_{I_{i}} \frac{\partial Q}{\partial x} \mathrm{~d} x=\frac{1}{\Delta x}\left(Q_{i+1 / 2}-Q_{i-1 / 2}\right) .
\end{equation}
Step 2: Through the finite volume update by using the cell averages and time-dependent flux at $t_n$, the cell averages at $t_*$ can be obtained.
From the temporal evolution for the interface values, we can obtain the conservative variables on the cell-interface at $t_*$.
Through the similar reconstruction strategy as shown in step 1, the time-dependent flux and temporal evolution for the interface values at $t_*$ can also be obtained.

Step 3:
Through the finite volume update by using the cell averages and time-dependent flux at $t_*$, we can obtain the cell averages at $t_{n+1}$.
Similarly, we can obtain the conservative variables in the cell-interface at $t_{n+1}$ through the temporal evolution model as $f^{n+1}=f^n+ \Delta t f^*_t$.

\subsection{\label{sec:stencils} The stencils for non-compact and compact reconstruction}
The stencils used for the 5th-order non-compact \cite{castro2011high} and compact reconstruction \cite{2004Hermite} are compared as shown in Fig. \ref{moban}. For the 5th-order non-compact reconstruction, in order to reconstruct the values on left and right interfaces of cell $I_i$, the five cell-averaged values in cells $I_{i-2}, I_{i-1}, I_{i}, I_{i+1}, I_{i+2}$ are used. While for the 5th-order compact reconstruction,  the cell-averaged values in cells $I_{i-1}, I_{i}, I_{i+1}$ and the cell-averaged derivatives in cells $I_{i-1}, I_{i+1}$ are used. To the same order, the stencils for compact reconstruction are smaller than non-compact reconstruction.

\section{\label{sec4} Accuracy test}

The three-dimensional advection of density perturbation is tested to verify the accuracy of HGKS with high-order non-compact reconstruction and compact reconstruction. The initial condition is given as follows:
\begin{equation}
\begin{aligned}
     \rho(x,y,z)=1+0.2sin(\pi(x+y+z)), U_i(x,y,z)=(1,1,1), p(x,y,z)=1.
\end{aligned}
\end{equation}
The computation domain is $[0,2]\times[0,2]\times[0,2]$ in three-dimension. 
In the computation, a series of uniform meshes with $N^3$ cells are used. The periodic
boundary condition is adopted in all directions. The analytic solution is:
\begin{equation}
\begin{aligned}
     \rho(x,y,z)=1+0.2sin(\pi(x+y+z-t)), U_i(x,y,z)=(1,1,1), p(x,y,z)=1.
\end{aligned}
\end{equation}

With the $r_n$th-order spatial reconstruction in normal directions, $r_t$th-order spatial reconstruction in tangential direction and S2O4 temporal discretization, the leading term of the truncation error \cite{ji2019high} is $O(\Delta x^{r_n}+\Delta y^{r_t}+\Delta z^{r_t}+\Delta t^4)$. 
The $L^1, L^2$ and $L^\infty$ errors and convergence orders at $t = 2$ are presented. 
To keep the $r$th-order accuracy in the test, $\Delta t=C\Delta x^{r/4}$ needs to be used for the $r$th-order scheme. To achieve a 2Mth- or (2M-1)th-order spatial accuracy, at least $M\times M$ Gaussian points are required for the cell interface \cite{ji2019high}.
The accuracy test for non-compact 5th-order scheme are shown in Table \ref{ac_n5_4g}.
The accuracy test for non-compact 7th-order scheme with 4 and 16 Gaussian points are shown in Table \ref{ac_n7_4g} and \ref{ac_n7_16g} respectively. It can be observed that when 4 Gaussian points are used, the accuracy of 7th-order cannot be maintained, and the accuracy will fall to 4th-order during mesh refinement. When it is increased to 16 Gaussian points, the numerical scheme can maintain 7th-order accuracy. Therefore, for high-order schemes, sufficient Gaussian points are needed to maintain high-order accuracy.
However, when 16 Gaussian points used, the amount of calculation is almost twice that of 4 Gaussian points. 
Specifically, in the efficiency test for the 7th-order non-compact scheme, in which the grids are set as $64 \times 96 \times 64$ in streamwise, normal-boundary and spanwise directions respectively for channel flow simulation, and 16 cores are used for parallel computation test, the central processing unit (CPU) time (s/step) for 4 Gaussian points is 0.961, while for 16 Gaussian points is 1.964. Therefore, for turbulence simulation, in order to save computing resources, we use 4 Gaussian points in the cell interface, but the results are still improved as shown in the following section.
 The accuracy test for the compact 5th-order scheme is shown in Table \ref{ac_c5} and the compact GKS can also achieve the designed accuracy. 

 \begin{table}[!h]
	\caption{\label{ac_n5_4g} Three-dimensional accuracy test: errors and convergence orders of 5th-order non-compact linear scheme with 4 Gaussian points with $\Delta t = 0.3\Delta x ^{1.25}$   }
	\vspace{3mm}
	\centering
	\begin{tabular}{c cc cc cc}
		\hline
		\hline
		Mesh number  & $L^1$ error&  order &  $L^2$ error&  order   & $L^{\infty}$ error&  order\\
		\hline
		$5^3 $       &7.7945E-02   &         &8.5715E-02    &       &1.2043E-01 &             \\
		$10^3$       &2.2295E-03   &5.13     &2.5293E-03    &5.08   &3.4448E-03 &  5.13        \\
		$20^3$       &6.1034E-05   &5.19     &6.8179E-05    &5.21   &9.6405E-05 &  5.16       \\
	$	40^3$       &1.8262E-06   &5.06     &2.0315E-06    &5.07   &2.8729E-06 &  5.07      \\

		\hline
		\hline
	\end{tabular}	

\vspace{-2mm}
\end{table}

 \begin{table}[!h]
	\caption{\label{ac_n7_4g} Three-dimensional accuracy test: errors and convergence orders of 7th-order non-compact linear scheme with 4 Gaussian points with $\Delta t = 0.3\Delta x ^{1.75}$   }
	\vspace{3mm}
	\centering
	\begin{tabular}{c ccc ccc ccc}
		\hline
		\hline
		Mesh number  & $L^1$ error&  order &  $L^2$ error&  order   & $L^{\infty}$ error&  order\\
		\hline
		$5^3$        &2.2004E-02   &         &2.4067E-02    &       &3.3999E-02 &             \\
		$10^3 $     &2.8164E-04   &6.29     &3.1237E-04    &6.27   &4.3516E-04 &  6.29        \\
		$20^3 $     &1.1384E-05   &4.63     &1.2593E-05    &4.63   &1.7619E-05 &  4.63       \\
		$40^3 $      &6.8005E-07   &4.07     &7.5457E-07    &4.06   &1.0640E-06 &  4.05      \\

		\hline
		\hline
	\end{tabular}	

\vspace{-2mm}
\end{table}

 \begin{table}[!h]
	\caption{\label{ac_n7_16g} Three-dimensional accuracy test: errors and convergence orders of 7th-order non-compact linear scheme with 16 Gaussian points with $\Delta t = 0.3\Delta x ^{1.75}$   }
	\vspace{3mm}
	\centering
	\begin{tabular}{c cc cc cc}
		\hline
		\hline
		Mesh number  & $L^1$ error&  order &  $L^2$ error&  order   & $L^{\infty}$ error&  order\\
		\hline
		$5^3 $       &1.9758E-02   &         &2.1738E-02    &       &3.0528E-02 &             \\
		$10^3$       &1.2131E-04   &7.35     &1.3509E-04    &7.33   &1.8743E-04 &  7.35        \\
		$20^3$       &6.8894E-07   &7.46     &7.7024E-07    &7.45   &1.0892E-06 &  7.43       \\
		$40^3 $      &5.7266E-09   &6.91     &6.3566E-09    &6.92   &8.9783E-09 &  6.92      \\

		\hline
		\hline
	\end{tabular}	

\vspace{-2mm}
\end{table}

%
%

 \begin{table}[!h]
	\caption{\label{ac_c5} Three-dimensional accuracy test: errors and convergence orders of 5th-order compact linear scheme with 4 Gaussian points with $\Delta t = 0.3\Delta x ^{1.25}$   }
	\vspace{3mm}
	\centering
	\begin{tabular}{c cc cc cc}
		\hline
		\hline
		Mesh number  & $L^1$ error&  order &  $L^2$ error&  order   & $L^{\infty}$ error&  order\\
		\hline
		$5^3 $       &5.8230E-02   &         &6.6867E-02    &       &8.9970E-02 &             \\
		$10^3$       &1.9314E-03   &4.91     &2.1135E-03    &4.98   &2.9841E-03 &  4.91        \\
		$20^3$       &5.5572E-05   &5.12     &6.1476E-05    &5.10   &8.6024E-05 &  5.12       \\
	$	40^3$        &1.7123E-06   &5.02     &1.9003E-06    &5.01   &2.6829E-06 &  5.00      \\

		\hline
		\hline
	\end{tabular}	

\vspace{-2mm}
\end{table}

\section{\label{sec5} ILES for turbulence simulation}
In this section, we apply the high-order non-compact scheme and compact scheme for turbulence simulation to investigate the effect by improving the order of the GKS and the compact reconstruction. Additionally, we also study the contribution of tangential flux to turbulence simulation in GKS. So in the simulation, we compare the results between the schemes with improvement of accuracy in the normal direction only and in both normal and tangential directions. In the following sections, the abbreviation “N7T5” indicates the scheme with 7th-order non-compact reconstruction in normal direction and 5th-order non-compact reconstruction in tangential direction, and the abbreviation “C5T5” indicates the scheme with 5th-order compact reconstruction in normal direction and 5th-order non-compact  reconstruction in tangential direction.

\subsection{Taylor-Green vortex}
Taylor-Green vortex is a classical problem and has been widely studied \cite{brachet1983small, gallis2017molecular, debonis2013solutions, bull2015simulation}.
The ILES of three-dimensional Taylor-Green vortex is conducted.
The flow is computed within a periodic square box defined as $-\pi L\leq x, y, z\leq \pi L$. With a uniform temperature, the initial condition is given by \cite{debonis2013solutions}
\begin{equation}
\begin{aligned}
U_1 = & V_0\sin(\frac{x}{L})\cos(\frac{y}{L})\cos(\frac{z}{L}),\\
U_2 = & -V_0\cos(\frac{x}{L})\sin(\frac{y}{L})\cos(\frac{z}{L}),\\
U_3 = & 0,\\
p = & p_0+\frac{\rho_0V_0^2}{16}(\cos(\frac{2x}{L})+\cos(\frac{2y}{L}))(\cos(\frac{2z}{L})+2).
\end{aligned}
\end{equation}
In the computation, $L=1, V_0=1, \rho_0=1$, and the Mach number takes $M_0=V_0/c_0=0.1$, where $c_0$ is the sound speed. The fluid is a perfect gas with $\gamma=1.4$, Prandtl number is $Pr=1$, and Reynolds number $Re=1600$. The characteristic convective time $t_c = L/V_0$.
In the computation, the grids of $256^3$ are used, which can not only get satisfactory numerical results but also show the differences between different schemes. 

Several statistic quantities are computed from the flow as it evolves in time.
The volume-averaged kinetic energy $E_k$ is defined by
\begin{equation}
\begin{aligned}
E_k=\frac{1}{\rho_0\Omega}\int_\Omega\frac{1}{2}\rho U_i \cdot U_i \text{d} \Omega,
\end{aligned}
\end{equation}
where $\Omega$ is the volume of the computational domain. 
The dissipation rate of kinetic energy $\varepsilon(E_k)$ is defined as
\begin{equation}
\begin{aligned}
\quad\varepsilon(E_k)=-\frac{\text{d}E_k}{\text{d}t}.
\end{aligned}
\end{equation}
For the incompressible limit, the dissipation rate is related to the integrated enstrophy \cite{cao2021high}
\begin{equation}
\begin{aligned}
\varepsilon(\zeta)=2\frac{\mu}{\rho_0}\zeta, \quad
&\zeta = \frac{1}{\rho_0\Omega}\int_\Omega\frac{1}{2}\rho \omega_i \cdot \omega_i \text{d} \Omega,
\end{aligned}
\end{equation}
where vorticity is $\omega_i=\varepsilon_{ijk} U_{k,j}$, $\varepsilon_{ijk}$ is the alternating tensor and $U_{k,j} = \partial U_i/\partial x_j$.

\begin{figure*} [!h]
\centering
    \subfigure[]{
        \includegraphics[height=0.4\textwidth]{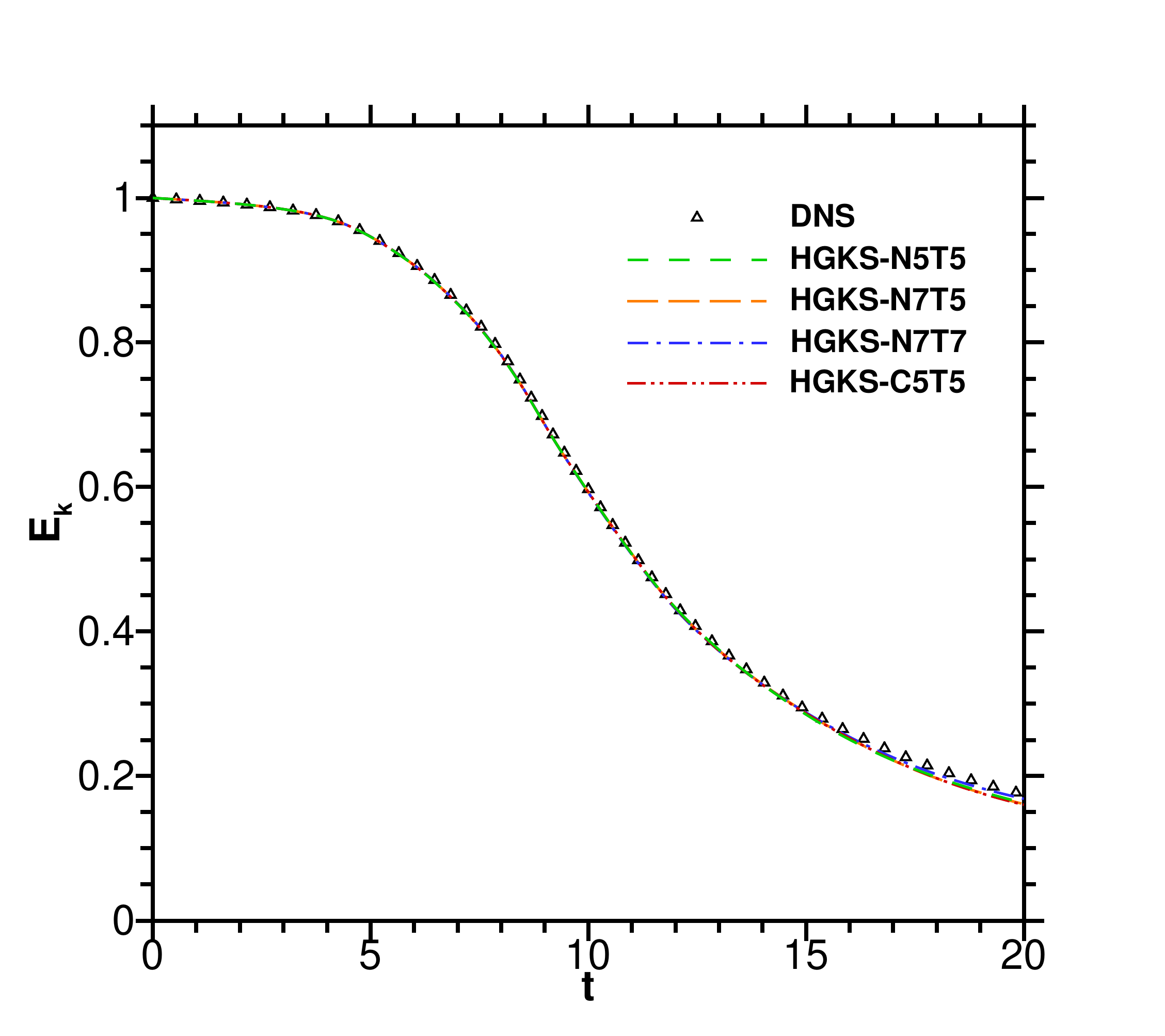}
    }
    \subfigure[]{
        \includegraphics[height=0.4\textwidth]{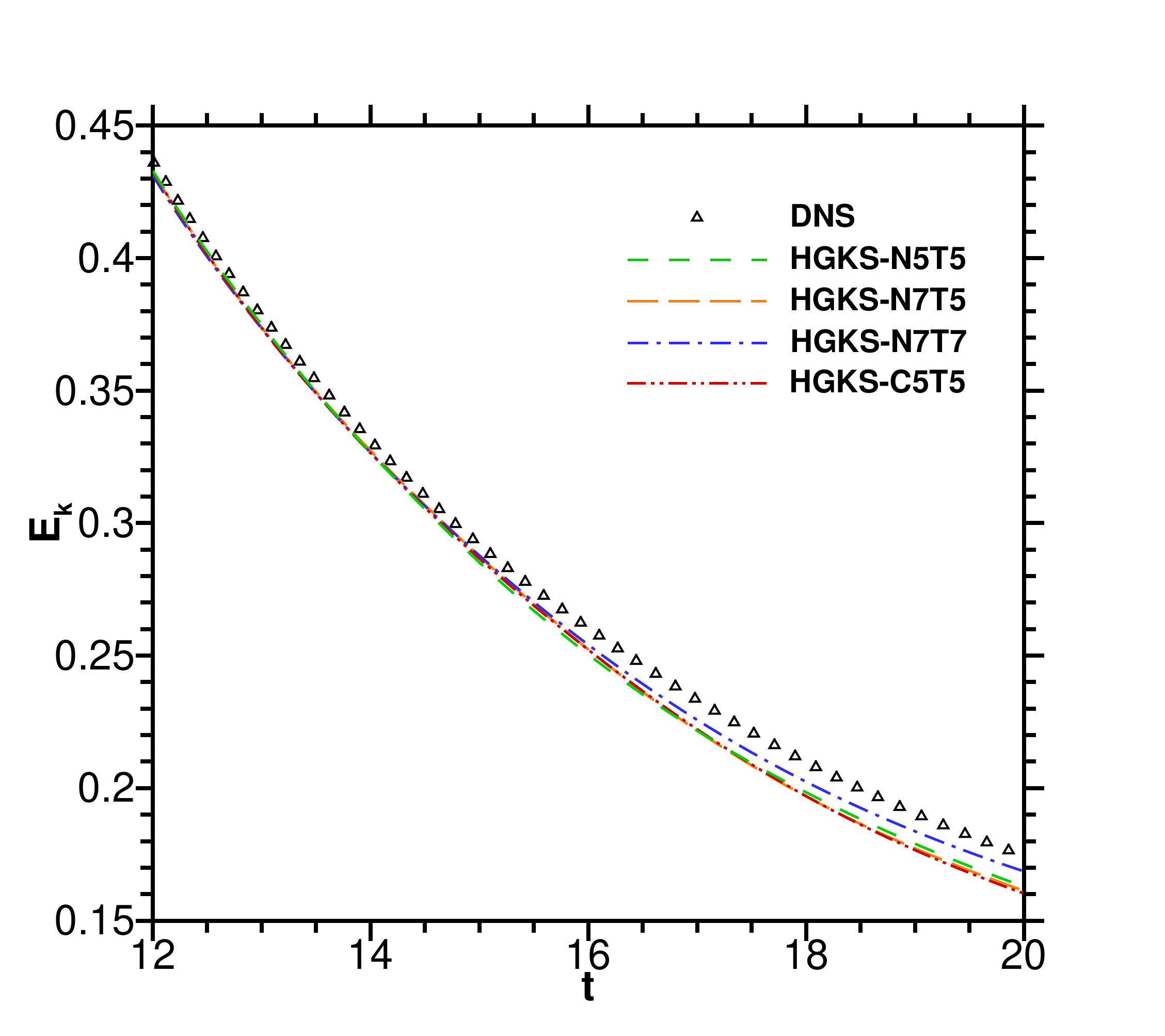}
    }   
	\caption{\label{tek_c} The time history of kinetic energy $E_k$ and local enlargement}			

    \subfigure[]{
        \includegraphics[height=0.4\textwidth]{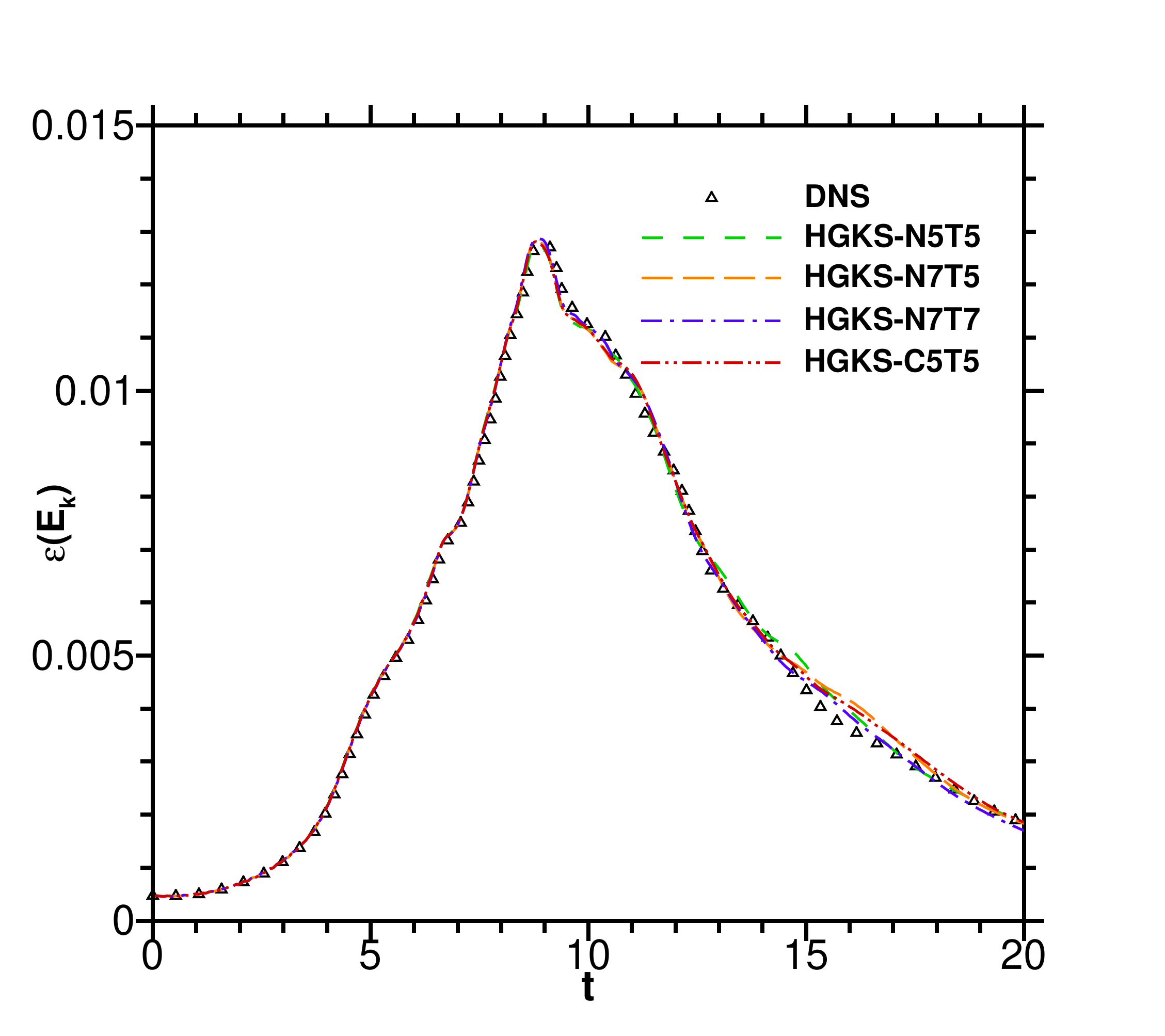}
    }
    \subfigure[]{
        \includegraphics[height=0.4\textwidth]{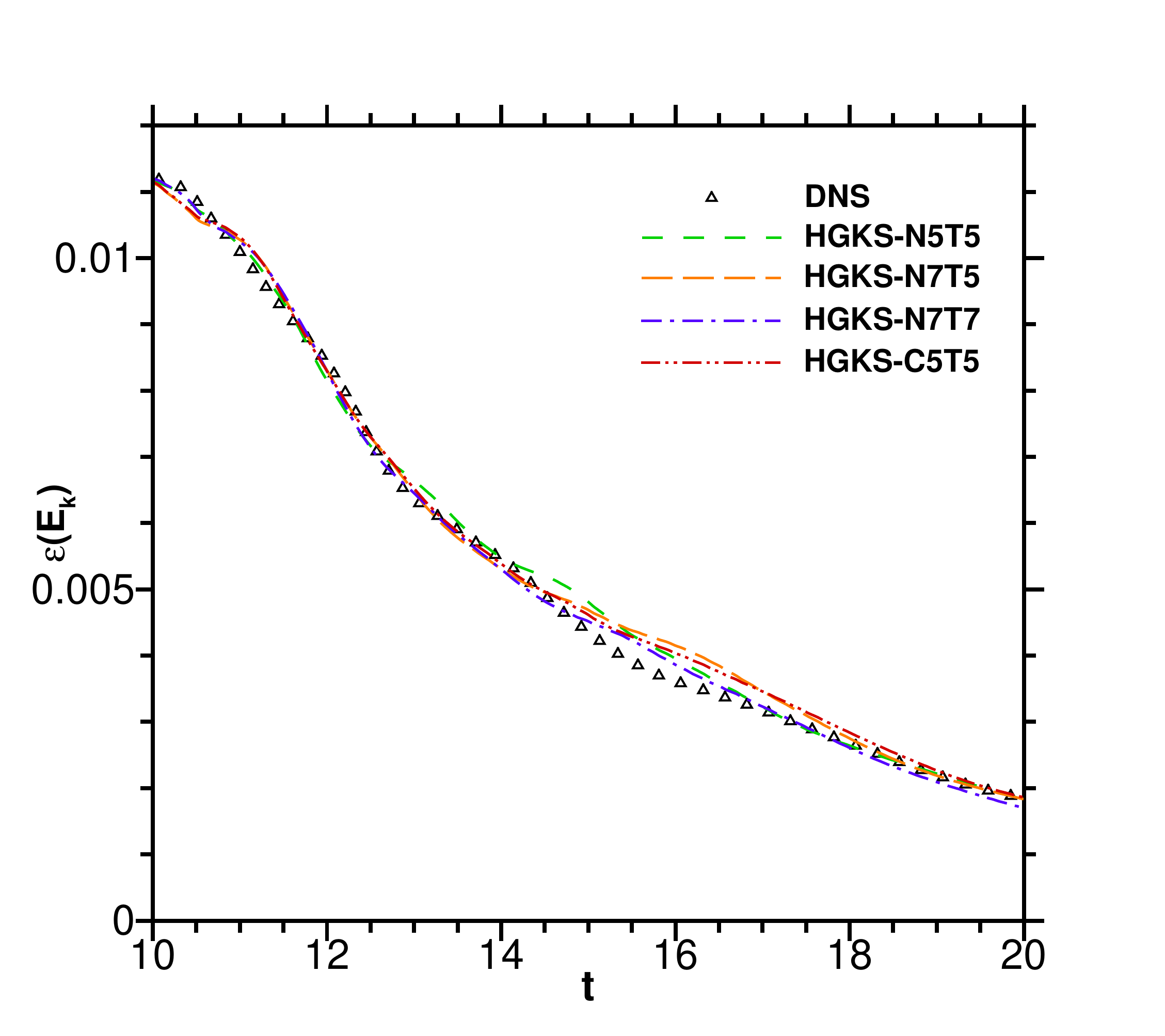}
    }   
	\caption{\label{disrate_c} The time history of dissipation rate $\varepsilon(E_k)$ and local enlargement}			
\end{figure*}

\begin{figure*}  [!h]
\centering
    \subfigure[]{
        \includegraphics[height=0.4\textwidth]{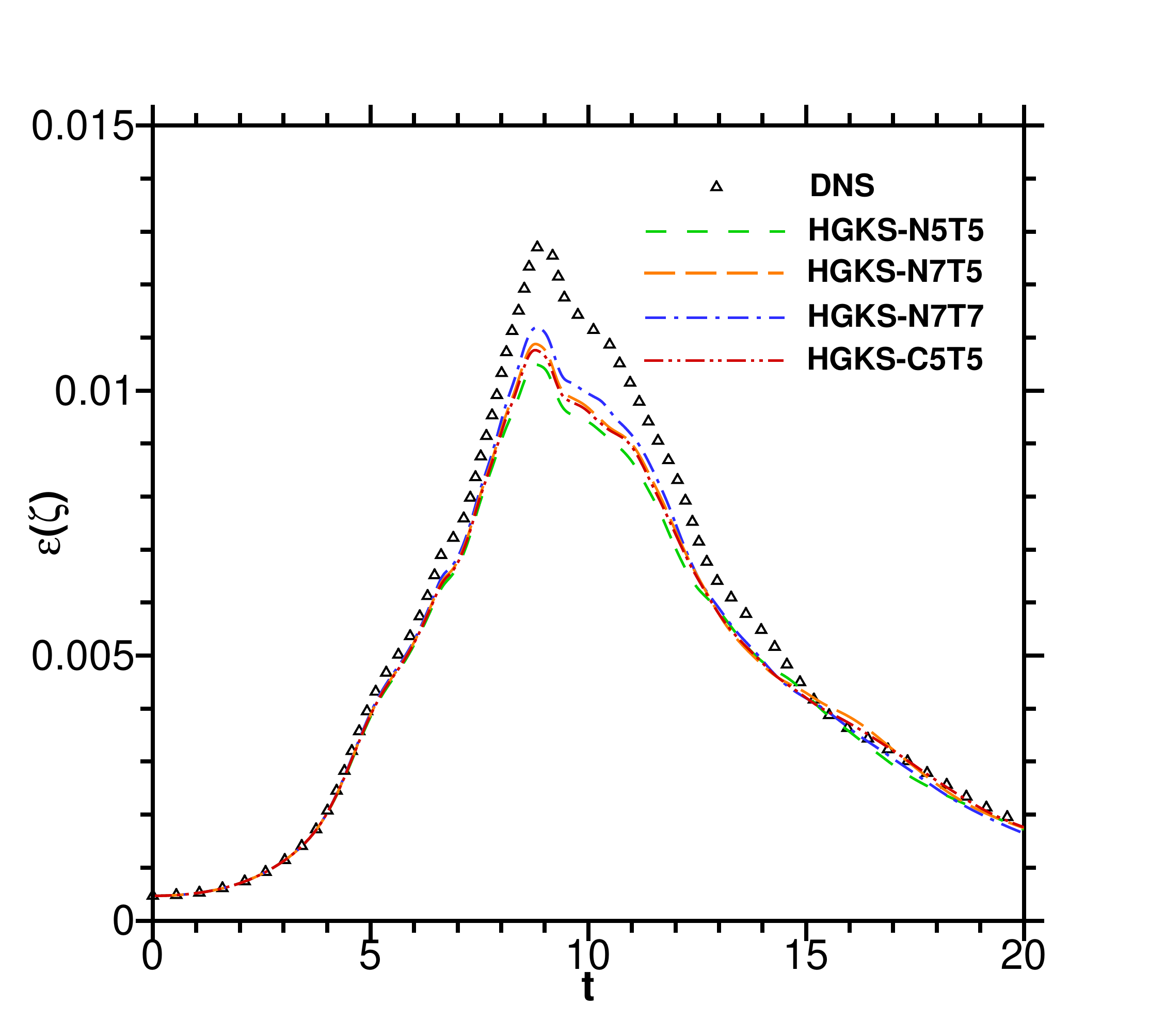}
    }
    \subfigure[]{
        \includegraphics[height=0.4\textwidth]{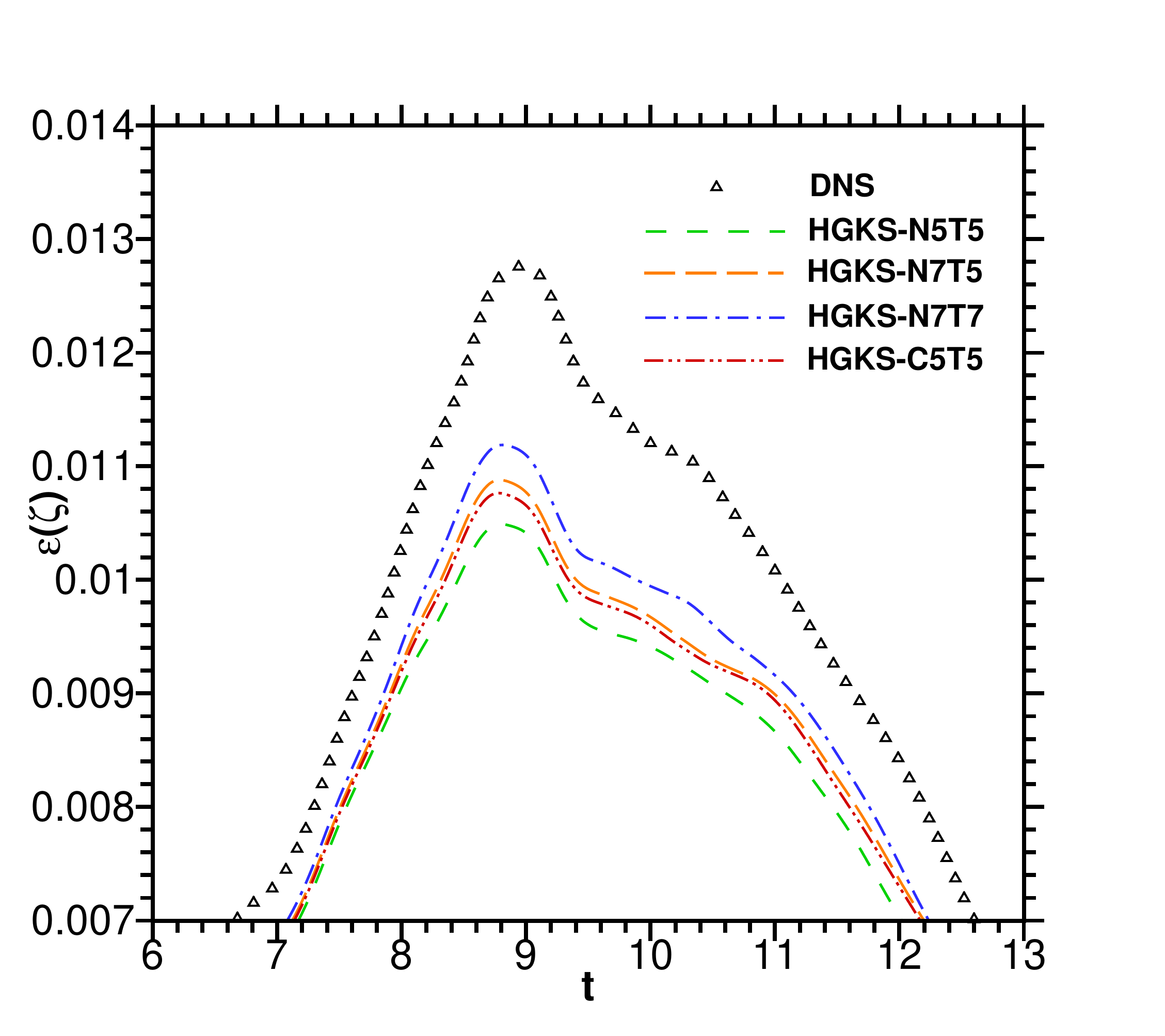}
    }   
	\caption{\label{enst_c} The time history of enstrophy $\varepsilon(\zeta)$ and local enlargement}

%
    \subfigure[]{
        \includegraphics[height=0.4\textwidth]{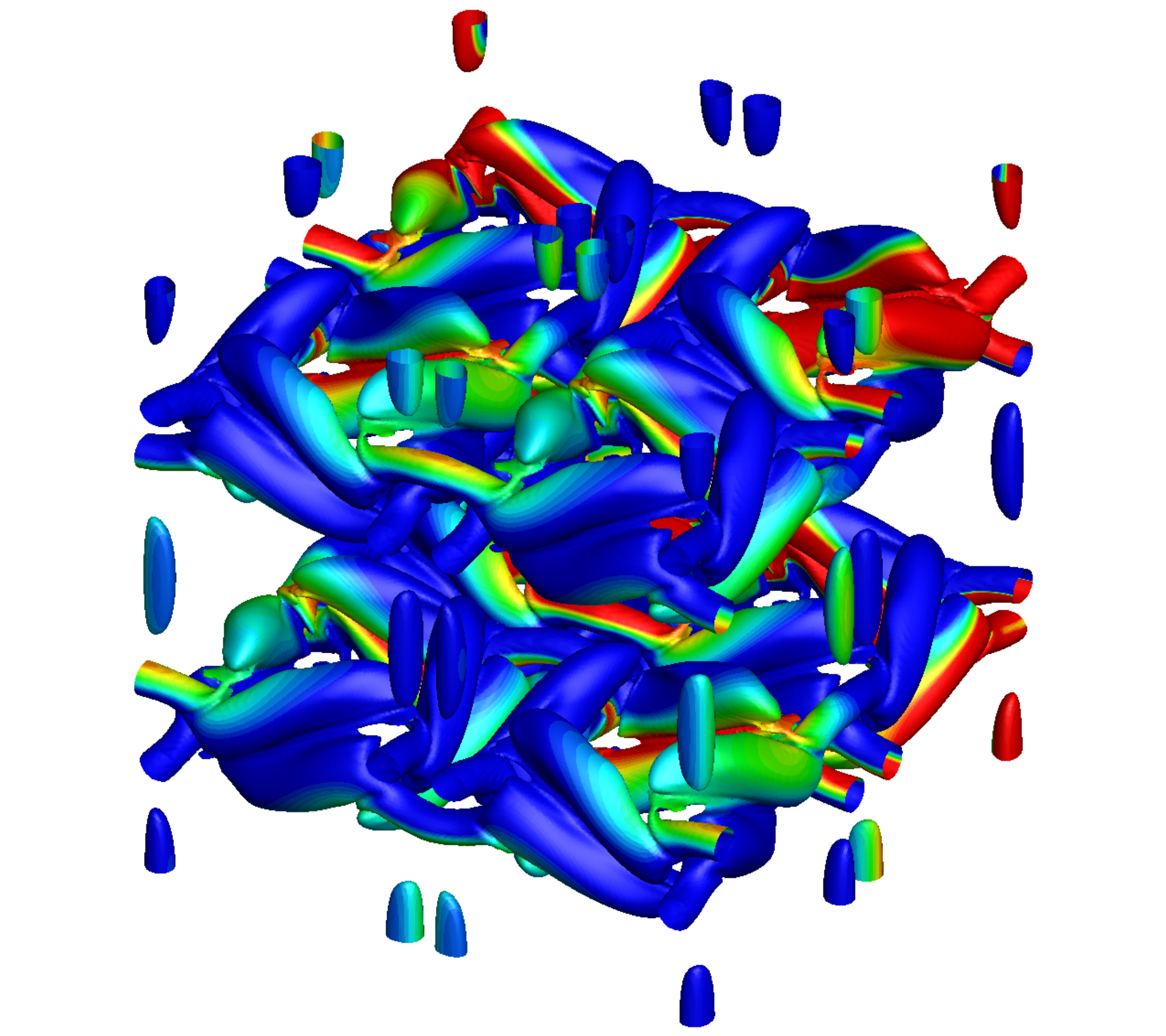}
    }
    \subfigure[]{
        \includegraphics[height=0.4\textwidth]{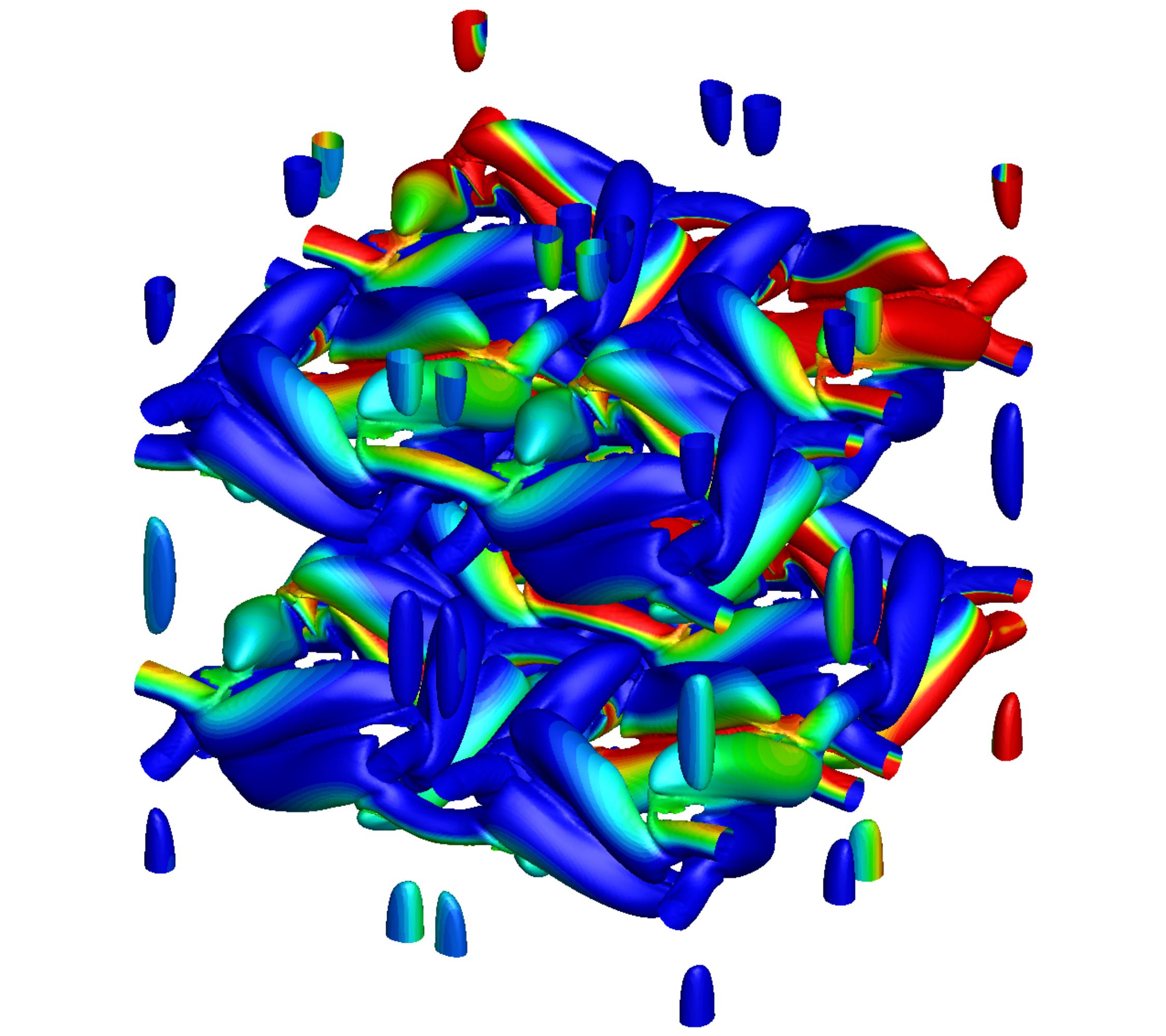}
    }   
	\caption{\label{tgv_qtu} The iso-surface of the second invariant of velocity gradient tensor $Q$ colored by velocity magnitude at t = 5 for (a) HGKS-N5T5 and  (b) HGKS-N7T7 }

\end{figure*}

Fig. \ref{tek_c} shows the time history of average kinetic energy and the local enlargement. Except for the last period time, ILES results are basically consistent with DNS. In the last period time, the results of “HGKS-N7T7” are better than other schemes. 
So the high-order reconstruction in the tangential direction is also important for the present ILES.
The kinetic energy dissipation rates $\varepsilon(E_k)$ and the local enlargement are shown in Fig. \ref{disrate_c}. The behaviors of the schemes in this statistics are similar to the time history of average kinetic energy. Except for the last period time, the results of the various schemes can match with DNS. In the last period time, the results of “HGKS-N7T7” are slightly closer to DNS than other schemes.
The enstrophy integral $\varepsilon(\zeta)$ and local enlargement are shown in Fig. \ref{enst_c}. It can be observed that both the higher-order reconstruction and compact reconstruction can improve the numerical results. Increasing the  reconstruction order in the tangential direction can also improve the results.
Hunt $et$ al.\cite{hunt1988eddies} identified vorticity of an incompressible flow as connected fluid regions with a positive second invariant of the velocity-gradient tensor as $Q=(U_{i,j}^2-U_{i,j}U_{j,i})/2$.
Q-criterion is an indication of vorticity prevailing overstrain and is helpful in identifying vortex cores. The Q-criterion iso-surfaces show the ability of the different schemes to resolve turbulent structures qualitatively.
The iso-surface of the second invariant of velocity gradient tensor $Q$ colored
by velocity magnitude at t = 5 for HGKS-N5T5 and HGKS-N7T7 are shown in Fig. \ref{tgv_qtu}. Velocity magnitude ranges from 0 to 0.2 and 20 equivalent levels are used. For the iso-surface of $Q$ in TGV case, the difference between different schemes is negligibly small.
 
In the numerical simulation, the overall dissipative behavior is determined by both physical and numerical dissipation. For the current study, the quantitative study of numerical dissipation is presented as well. The kinetic energy dissipation rate obtained from the Navier-Stokes equations is the sum of three contributions \cite{cao2021highdns}:
\begin{equation}
\begin{aligned}
	&\varepsilon_1=\frac{2\mu}{\rho_0 \Omega}\int_\Omega
	S^{\ast}_{ij}:S^{\ast}_{ij}\text{d}\Omega,\\
	&\varepsilon_2=\frac{\mu_b}{\rho_0 \Omega}\int_\Omega
	\theta^2\text{d} \Omega,\\
	&\varepsilon_3=-\frac{1}{\rho_0\Omega}\int_\Omega
	p \theta \text{d} \Omega,
\end{aligned}
\end{equation}
where $S^{\ast}_{ij}$ is the deviatoric part of the strain rate
tensor $S_{ij}$, with $S^{\ast}_{ij} = S_{ij} - \delta_{ij} S_{kk}/3$, $S_{ij} = (U_{i,j} + U_{j,i})/2$.
The operator $(:)$ denotes the product for second-order tensor, 
and $\mu_b$ is the bulk viscosity. In current HGKS, the inherent bulk viscosity reads $\mu_b=\frac{2N}{3(N+3)}\mu$, where $N=2$ for the diatonic gas. $\theta = U_{i,i}$ denotes the divergence of turbulent velocity. To suppress the error from numerical discretization, all spatial derivatives are computed by sixth-order central difference for three components of dissipation rate. 
Therefore, the numerical dissipation can be quantitatively computed by
\begin{align}
\varepsilon_{\text{num}}
=\varepsilon(E_k)-(\varepsilon_1+\varepsilon_2+\varepsilon_3).
\end{align}

\begin{figure*} 
    \subfigure[]{
        \includegraphics[height=0.4\textwidth]{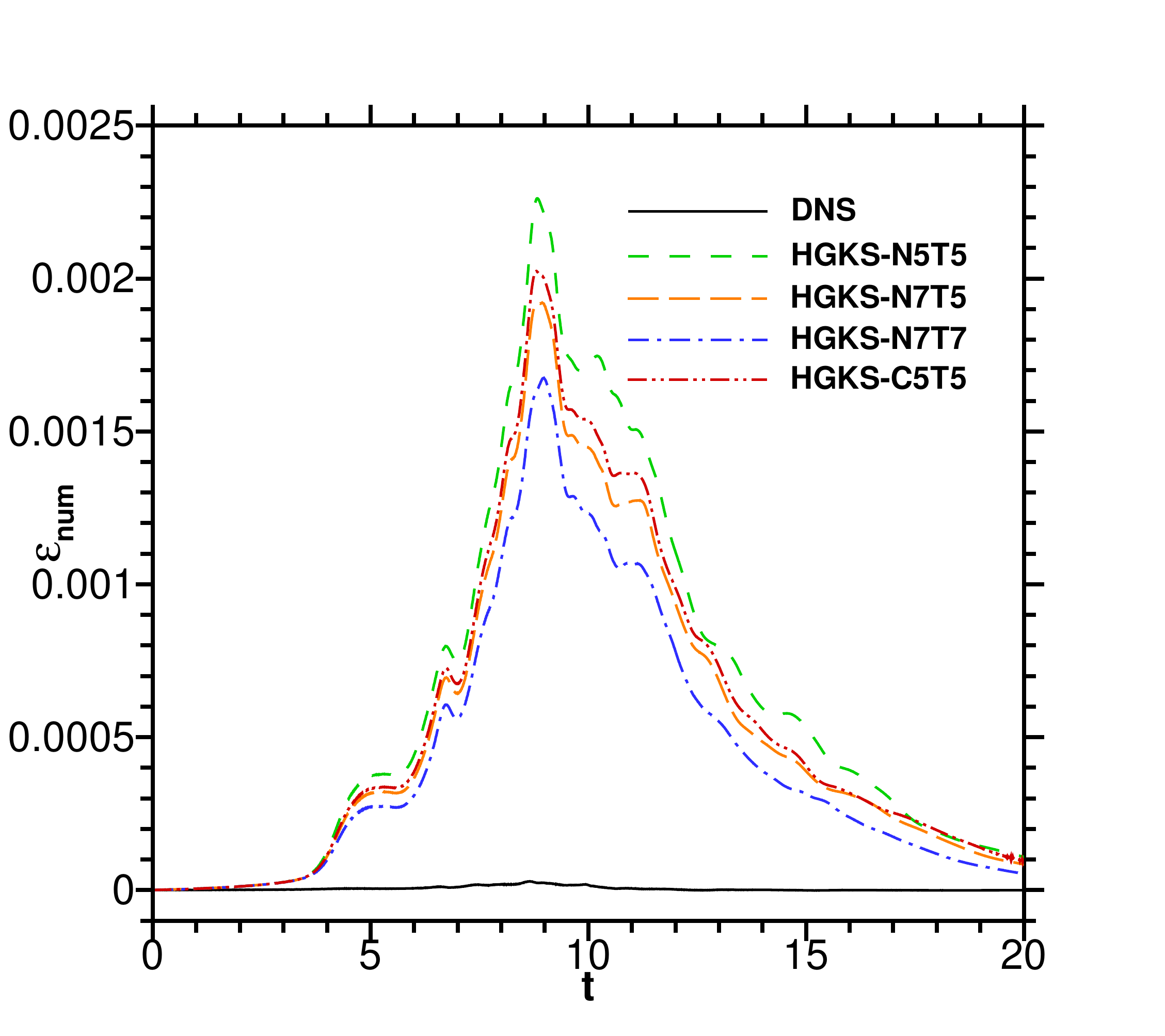}
    }
    \subfigure[]{
        \includegraphics[height=0.4\textwidth]{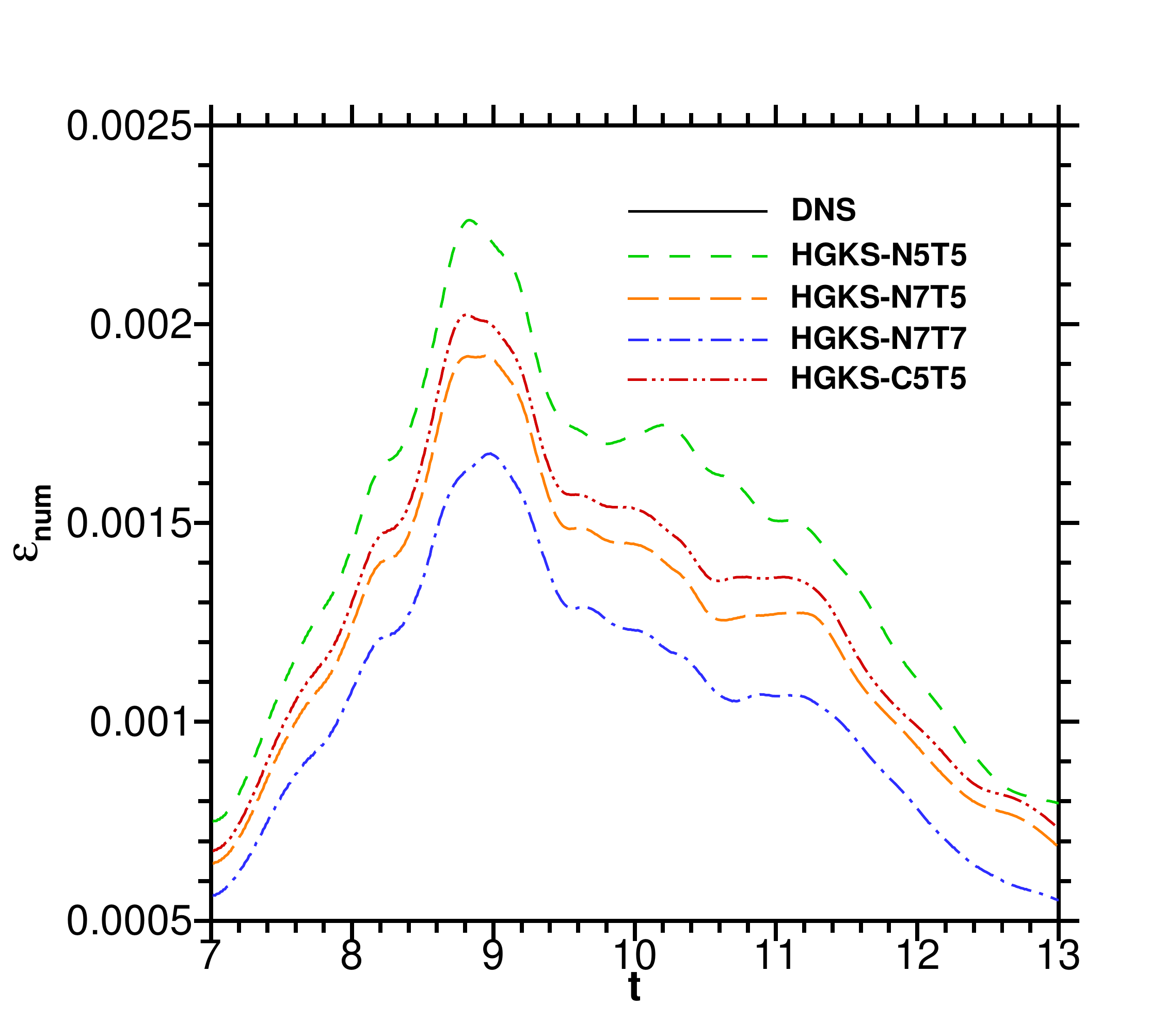}
    }   
\caption{ \label{disnum_c} The time history of numerical dissipation $\varepsilon_{num}$}

\end{figure*}

 The temporal evolution of numerical dissipation is shown in Fig. \ref{disnum_c}.
By using the higher-order reconstruction or compact reconstruction, the numerical dissipation during simulation can be reduced, and the corresponding ILES results are improved as shown before. This observation quantitatively illustrates the advantages of using higher-order or compact numerical schemes for ILES. The 5th-order scheme for ILES is still over-dissipative, and reducing the numerical dissipation can improve the results.

%
%

\subsection{Turbulent Channel flow}
Turbulent channel flow has been widely studied by numerical simulation \cite{kim1987turbulence, moser1999direct, kokkinakis2015implicit}. In this study, channel flow with friction Reynolds numbers of $Re_{\tau}=180$ and $ 395$ is numerically investigated to show the effect of using the higher-order scheme and the compact scheme.

The initial density and Mach number for the channel flow are set to $\rho=1$ and $Ma=0.1$, respectively. The non-dimensional domain size is set to $(x,y,z)\in[0,2\pi]\times[-1,1]\times[0,\pi]$ in the streamwise, normal-boundary, and spanwise directions, respectively \cite{zhao2021high}.
For the normal-boundary direction, the TANH function is used to generate the non-uniform grids, $i.e.$, 

\begin{equation}
\begin{aligned}
\displaystyle y=\tanh(b_g(\frac{\eta}{1.5\pi}-1))/\tanh(b_g)
\end{aligned}
\end{equation}
where $y\in[-1,1]$ and $\eta\in[0,3\pi]$. For $Re_{\tau}=180$, we set $b_g=2$. For $Re_{\tau}=395$, we set $b_g=2.5$ to refine the grids in near wall region and get more accurate velocity profile. In the streamwise and spanwise directions, periodic boundary conditions are used. In the normal-boundary direction, non-slip and isothermal boundary conditions are used, and the wall temperature is set as the initial fluid temperature.
For numerical simulation of turbulent channel flow at $Re_\tau=180$, the grids are set to be $96\times64\times64$ in normal-boundary, streamwise and spanwise directions respectively, as shown in Table \ref{g1}.

The initial streamwise velocity field is given by Poiseuille flow added with white noise as $ U(y) = 1.5(1-y^{2}) + \text{white noise}$. The white noise is set as $10\%$ amplitude of local streamwise velocity.
The setting of fluid viscosity is referred to previous study \cite{cao2021high} and is briefly illustrated below. The friction Reynolds number is defined as $Re_{\tau}=\rho U_{\tau}H/\mu$, where $H=1$, and the friction velocity $U_\tau$ is given by $U_{\tau}=\sqrt{\tau_{wall}/\rho}$, where $\tau_{wall}=\mu \frac{\partial U}{\partial y}|_{wall}$.
The logarithmic formulation is given by $U^+=\ln{Y^+}/\kappa+B$, where $\kappa=0.40$ and $B=5.5$ are selected for the low-Reynolds-number turbulent channel flow \cite{kim1987turbulence}. The normalized wall distance and normalized velocity are defined as $Y^+=\rho U_{\tau}y/\mu$ and $U^+=U/U_{\tau}$, respectively.
The plus-velocity at the channel centerline is estimated as 
	\begin{equation}
\begin{aligned}
			U_c^+=2.5\ln Y_c^++5.5
\end{aligned}
\end{equation}	
where $Y_c^+=180$ at the center line. 
The frictional velocity is determined from the relation 
			
	\begin{equation}
\begin{aligned}
			U_{\tau}=U_c/U_c^+
\end{aligned}
\end{equation}
where $U_c=1$ denotes the centerline velocity. The fluid viscosity is set to $\mu=\rho U_\tau H/Re_\tau=2.83\times10^{-4}$.

For the turbulent channel flow at $Re_\tau=395$, we use the same method as described above to determine the fluid viscosity and set $\mu=1.24\times10^{-4}$. We use $96^3$ grid points in the simulation, as shown in Table \ref{g1}.
During the simulation, the flow is driven by an external force, which maintains a constant flow flux in the streamwise direction \cite{kim1987turbulence}. After approximately 400 characteristic periodic time as 400 $H/U_c$, the laminar flow fields transit to turbulence. 

The efficiency for the non-compact scheme and compact scheme are investigated, in which 4 Gaussian point are used in cell-interface for both schemes. The central processing unit (CPU) time for different schemes are shown in Table \ref{cputime_compact}, in which the grids are set as $64 \times 96 \times 64$ in streamwise, normal-boundary and spanwise direction, respectively, and 16 CPU cores are used for parallel computation test. It can be found that using higher-order reconstruction will increase the amount of computation. Using compact reconstruction will also increase the amount of computation, but the increment is relatively small.

The flow statistics considered in this study are described below. For channel flow, the spatially average variable $\phi$ is calculated in the homogeneous directions, $i.e.$, the streamwise ($x$) and spanwise ($z$) directions as $<\phi>_{xz}$, where $<>_{xz}$ indicates the spatial averaging in the $x$-$z$ plane.
The mean velocity is calculated as $U_{ave}=<U>_{xz}$. The velocity fluctuating components $U_i'$ are calculated as $U'_i=U_i-<U_i>_{xz}$.

The normalized Reynolds stresses are defined as

\begin{equation}
\begin{aligned}
      RS(U'_iU'_j)=\frac {<U'_iU'_j>_{xz} } {(\overline U_\tau)^2},
\end{aligned}
\end{equation}
where $\overline U_\tau$ denotes the resolved friction velocity, which is used as the normalization factor.
The normalized root-mean-square of velocity fluctuation is defined as
\begin{equation}
\begin{aligned}
      RMS(U'_i)=\frac {  <U'^2_i>_{xz}^{\frac 12}  } {\overline U_\tau},
\end{aligned}
\end{equation}
The above-mentioned statistics are further averaged over the statistical time, and 200 $H/U_c$ is used as statistical time.
%

Tables \ref{re180} and \ref{re395_96} show the values of $U_\tau$ and $Re_\tau$ calculated from the simulated turbulent channel flows at $Re_\tau=180$ and $Re_\tau=395$, respectively.
 It is shown that, when using the higher-order reconstruction or compact reconstruction, the values of $Re_\tau$ are all improved.


%
%

 \begin{table}[!h]
	\caption{\label{g1} Grid setup for simulation.}
	\vspace{3mm}
	\centering
	\begin{tabular}{c cc c c}
		\hline
		\hline
		Case        & $N_y/\Delta y_{min}^+/\Delta y_{max}^+$      & $N_x/\Delta x^+$    &  $N_z/\Delta z^+$\\
		\hline

		G1/$Re_\tau=180$     &96/0.29/7.76   & 64/17.66    &64/8.33  \\
		G2/$Re_\tau=395$     &96/0.58/20.83   &96/25.84     &96/12.92 \\
  
		\hline
		\hline
	\end{tabular}	

\vspace{-2mm}
\end{table}

 \begin{table}[!h]
	\caption{\label{cputime_compact} The CPU time (s/step) for different schemes.}
	\vspace{3mm}
	\centering
\begin{tabular}{cccccc}
\hline
\hline
The scheme & HGKS-N5T5 &HGKS-N7T5&HGKS-N7T7& HGKS-C5T5 \\
  \hline
  The CPU time (s/step) & 0.595 & 0.649 & 0.961 &0.618 \\
\hline
\hline
\end{tabular}
\vspace{-2mm}
\end{table}

%
%


 \begin{table}[!h]
	\caption{\label{re180} Values of $U_\tau$ and $Re_\tau$ for $Re_\tau=180$ channel flow in $64\times96\times64$ grids.}
	\vspace{3mm}
	\centering
\begin{tabular}{cccccc}
\hline
\hline
The schemes &DNS & HGKS-N5T5 &HGKS-N7T5& HGKS-N7T7 & HGKS-C5T5 \\
\hline
 $U_\tau$  &-- & 0.0515     & 0.0532   & 0.0536   & 0.0530  \\
$Re_\tau$ & 178.12& 171.33  &177.04  & 178.44  &176.27\\
\hline
\hline
\end{tabular}
\vspace{-2mm}
\end{table}

%
%

 \begin{table}[!h]
	\caption{\label{re395_96} Values of $U_\tau$ and $Re_\tau$ for $Re_\tau=395$ channel flow in $96^3$ grids.}
	\vspace{3mm}
	\centering
\begin{tabular}{cccccc}
\hline
\hline
The schemes &DNS & HGKS-N5T5 &HGKS-N7T5& HGKS-N7T7 & HGKS-C5T5 \\
\hline
 $U_\tau$ &--     & 0.0446    & 0.0469    & 0.0471  & 0.0467    \\
$Re_\tau$ &392.24  &360.35   & 378.97  &380.45 &  377.05\\
\hline
\hline
\end{tabular}
\vspace{-2mm}
\end{table}
%
%

\subsubsection{Mean flow velocity profiles}
\begin{figure*} [!h]
\centering
    \subfigure[]{
        \includegraphics[height=0.4\textwidth]{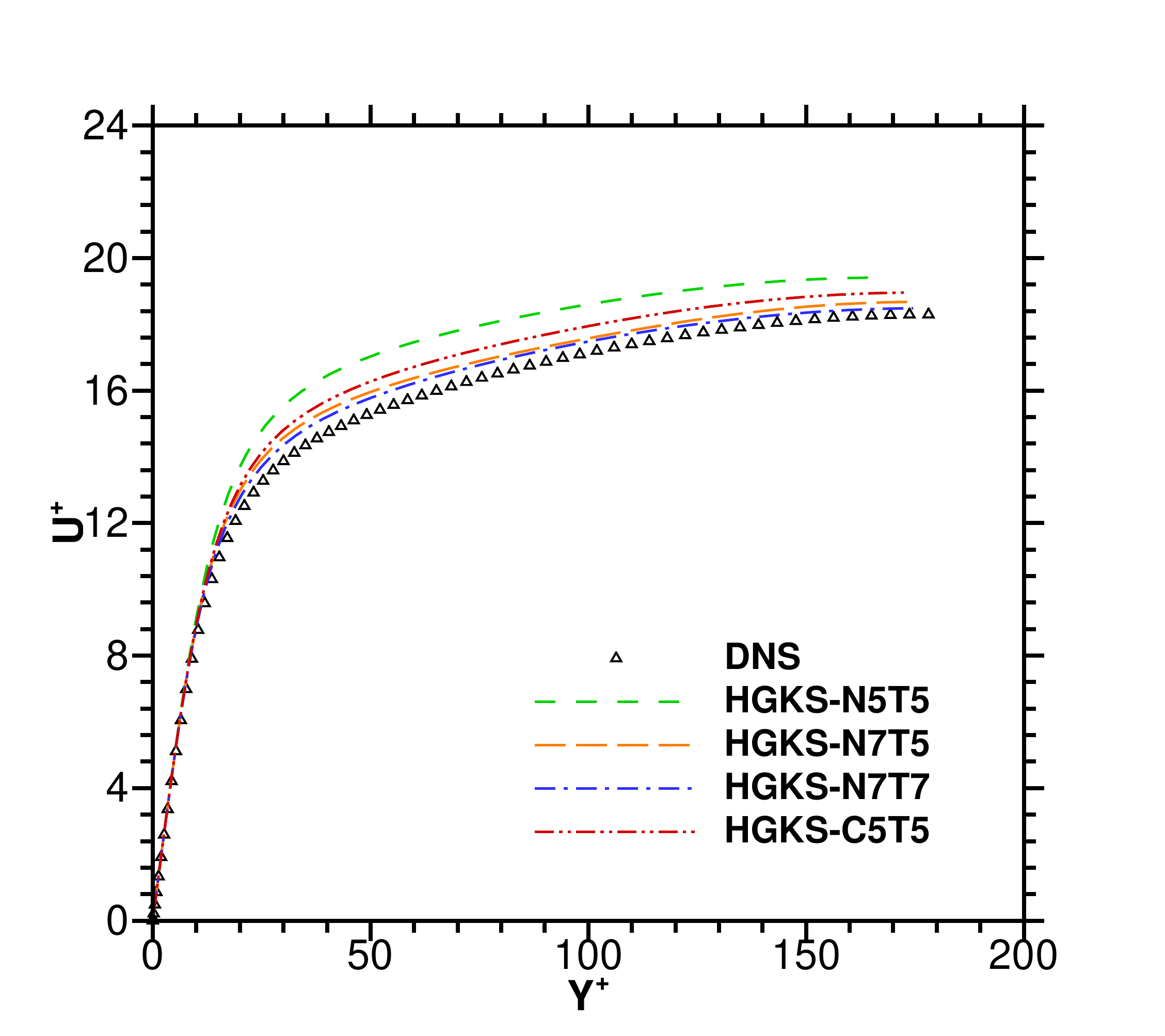}
    }
    \subfigure[]{
        \includegraphics[height=0.4\textwidth]{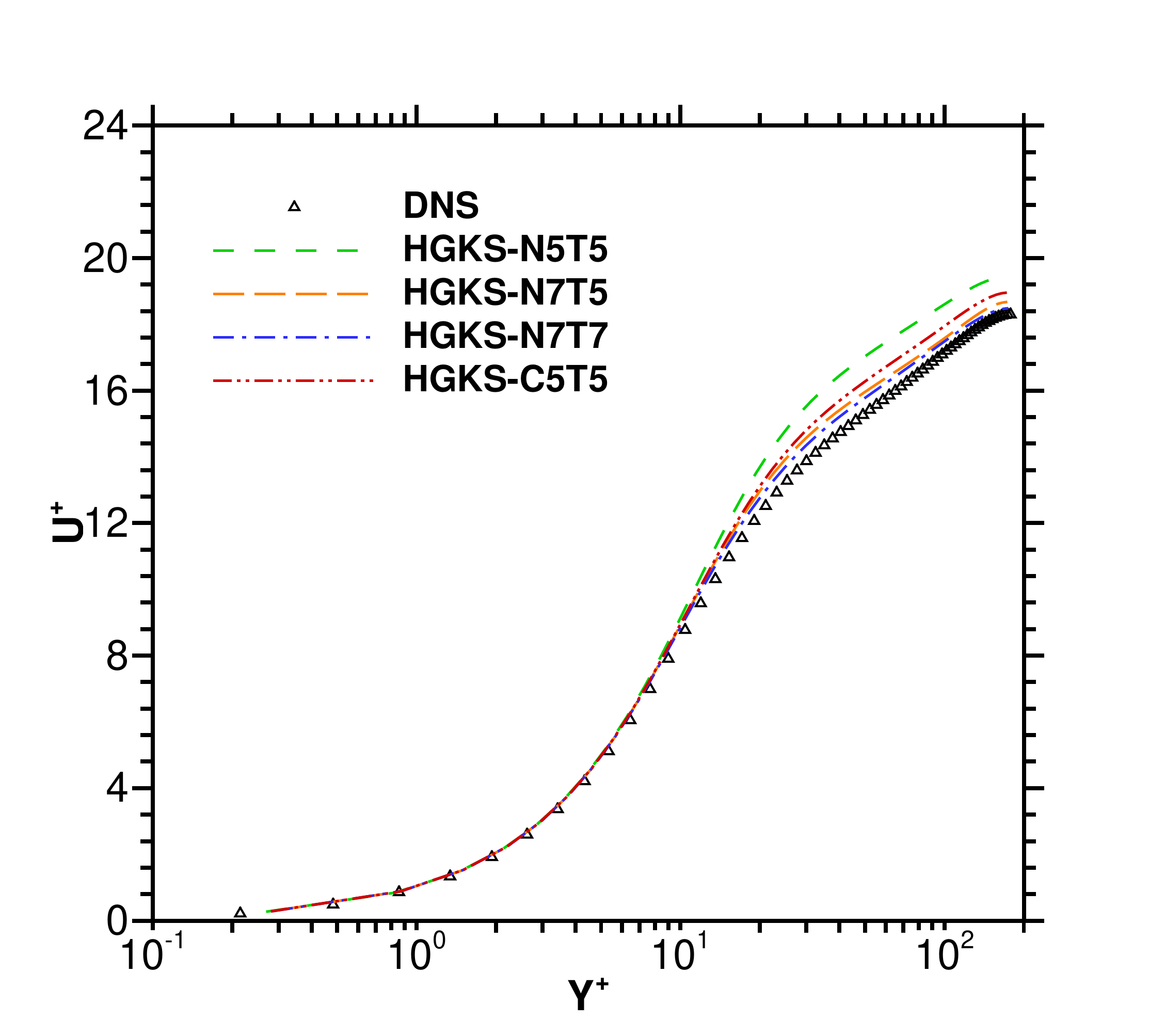}
    }   
\caption{\label{180_uave} Mean velocity profiles normalized by $U_\tau$ for $Re_\tau=180$ on (a) linear--linear and (b) log--linear plots.}	
%
%
    \subfigure[]{
        \includegraphics[height=0.4\textwidth]{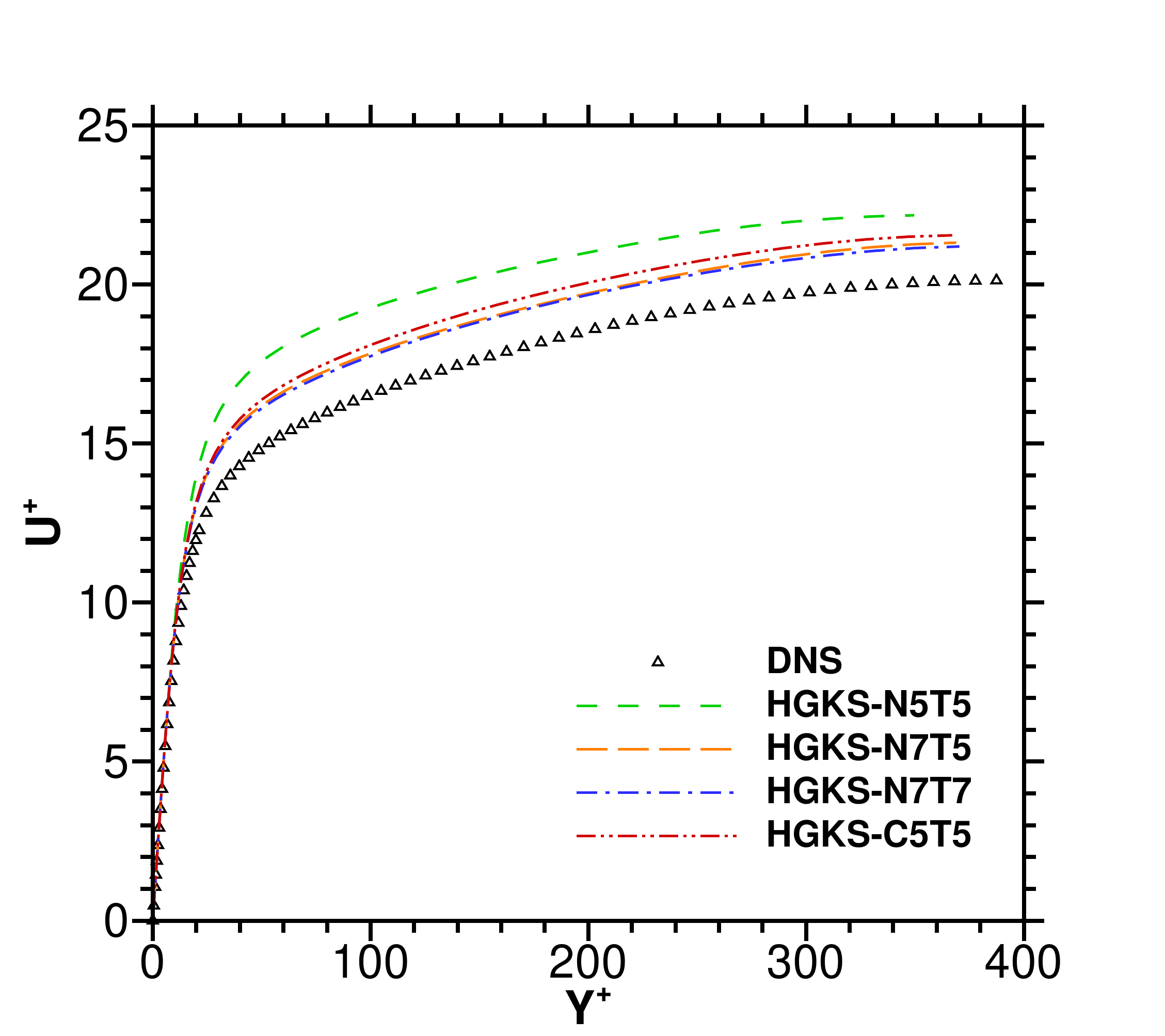}
    }
    \subfigure[]{
        \includegraphics[height=0.4\textwidth]{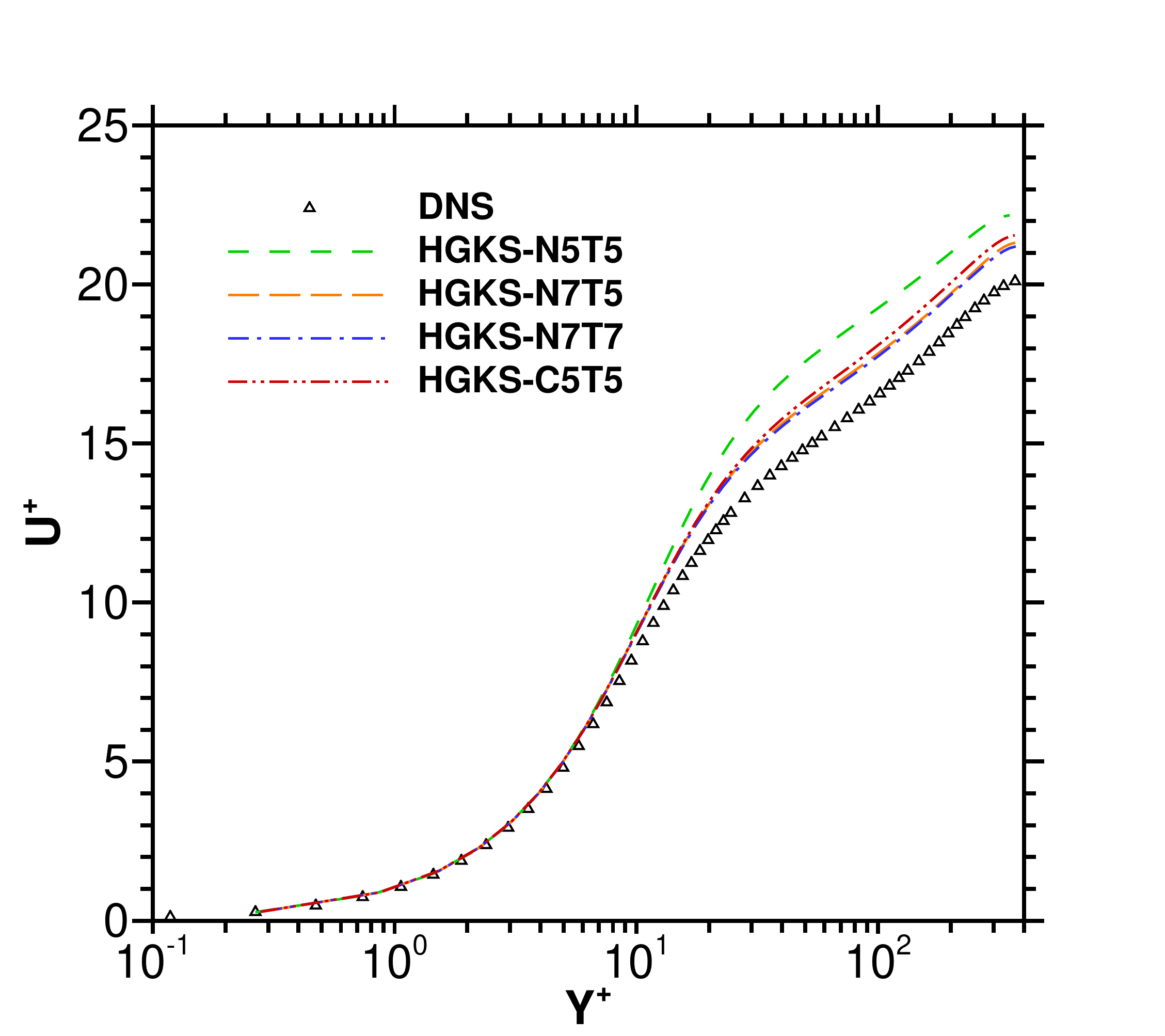}
    }   
\caption{\label{395_uave} Mean velocity profiles normalized by $U_\tau$ for $Re_\tau=395$ on (a) linear--linear and (b) log--linear plots.}	
\end{figure*}

The mean flow velocity profiles normalized by the resolved value of $U_\tau$ for the channel flow at $Re_\tau = 180$ is shown in Fig. \ref{180_uave}. All reference data are from DNS solution of incompressible turbulent channel flow by Moser $et$ $  al$\cite{moser1999direct}. The results from the 7th-order scheme are closer to DNS, as compared to the 5th-order scheme. Furthermore, improving the order in reconstruction from 5th to 7th in the tangential direction can further improve the results, especially in the near center-line region.
The results from 5th-order compact scheme are closer to DNS than that from the 5th-order non-compact scheme.

The mean flow velocity profiles normalized by the resolved value of $U_\tau$ for the channel flow at $Re_\tau = 395$  is shown is Fig. \ref{395_uave}. It is shown again that increasing the order of the scheme or using the compact scheme all can  significantly improve the numerical results. 

\subsubsection{Turbulence intensities}
\begin{figure*} [!h]
\centering
    \subfigure[]{
        \includegraphics[height=0.4\textwidth]{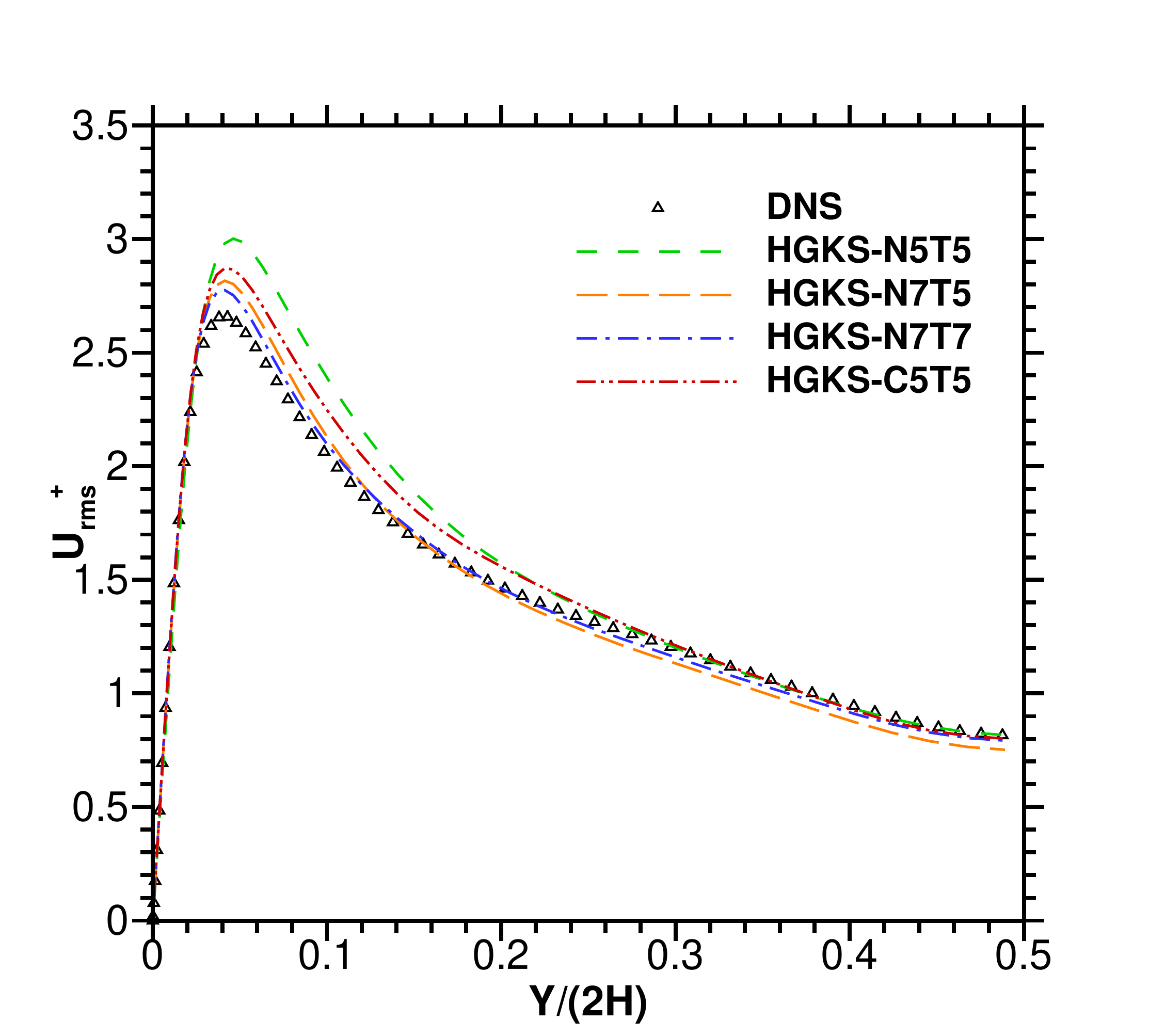}
    }
    \subfigure[]{
        \includegraphics[height=0.4\textwidth]{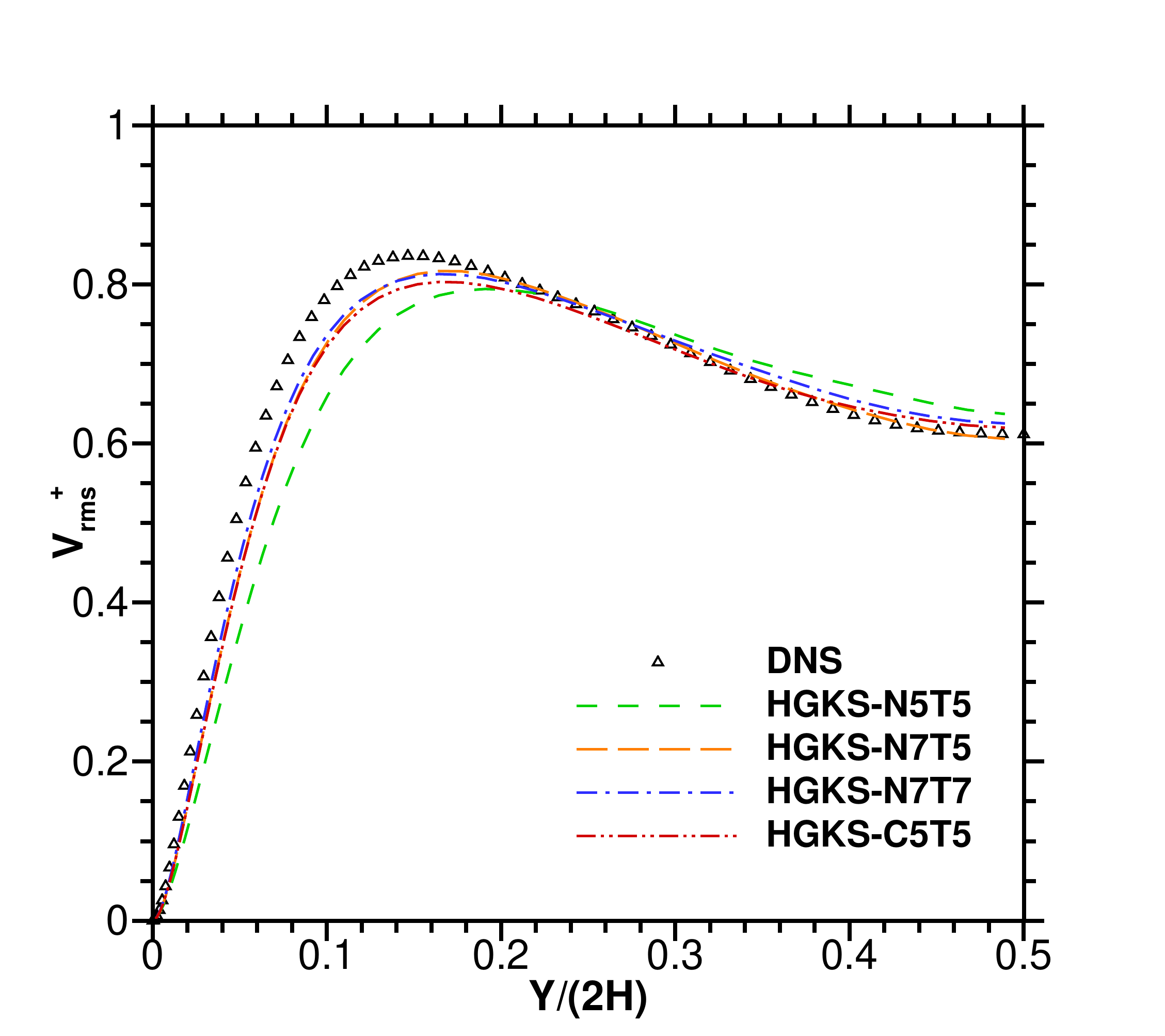}
    }   
	\caption{\label{180_urms} Root-mean-square fluctuation velocity profiles in (a) streamwise and (b) normal-boundary directions for $Re_\tau=180$.}	

%
%

    \subfigure[]{
        \includegraphics[height=0.4\textwidth]{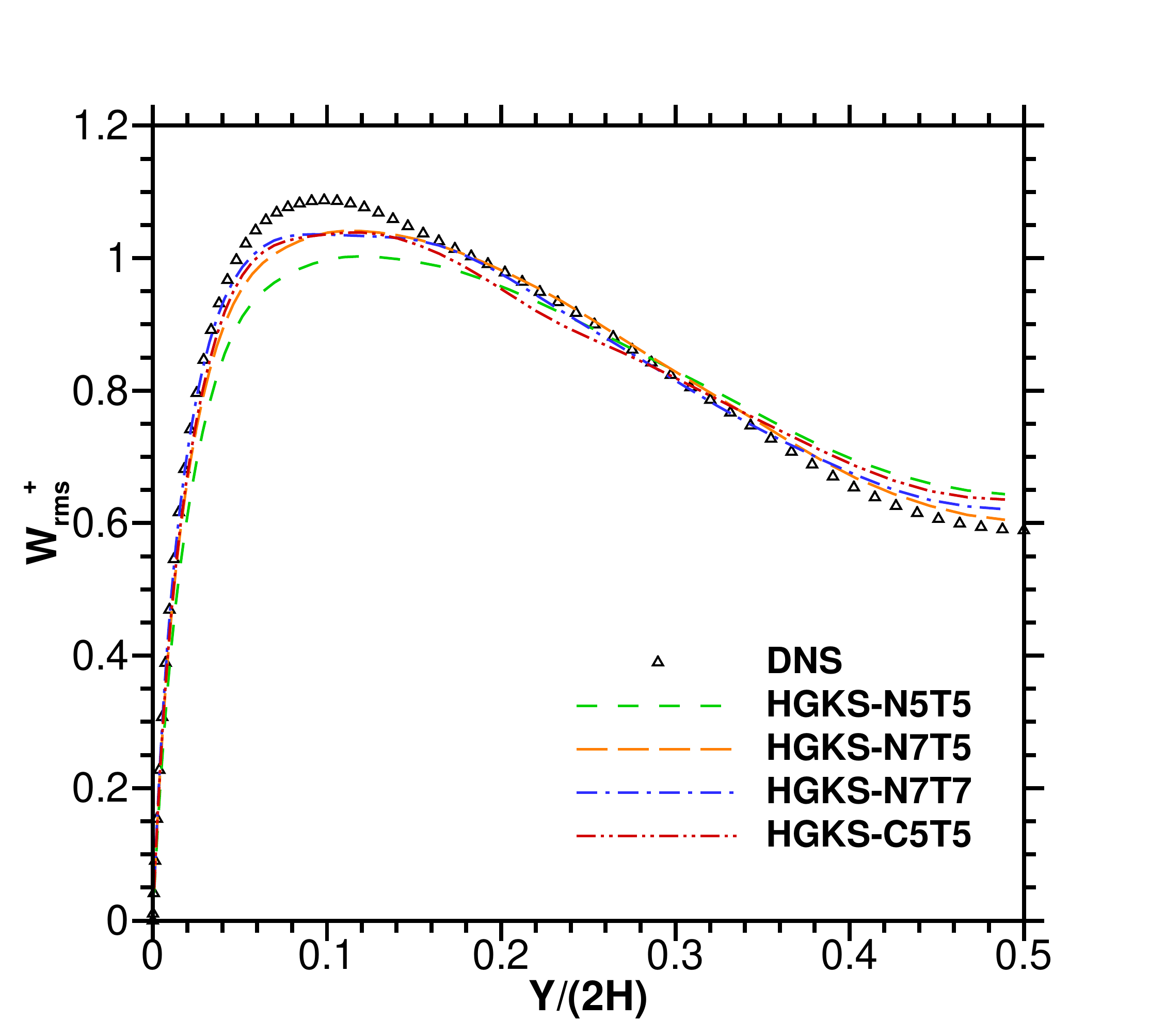}
    }
    \subfigure[]{
        \includegraphics[height=0.4\textwidth]{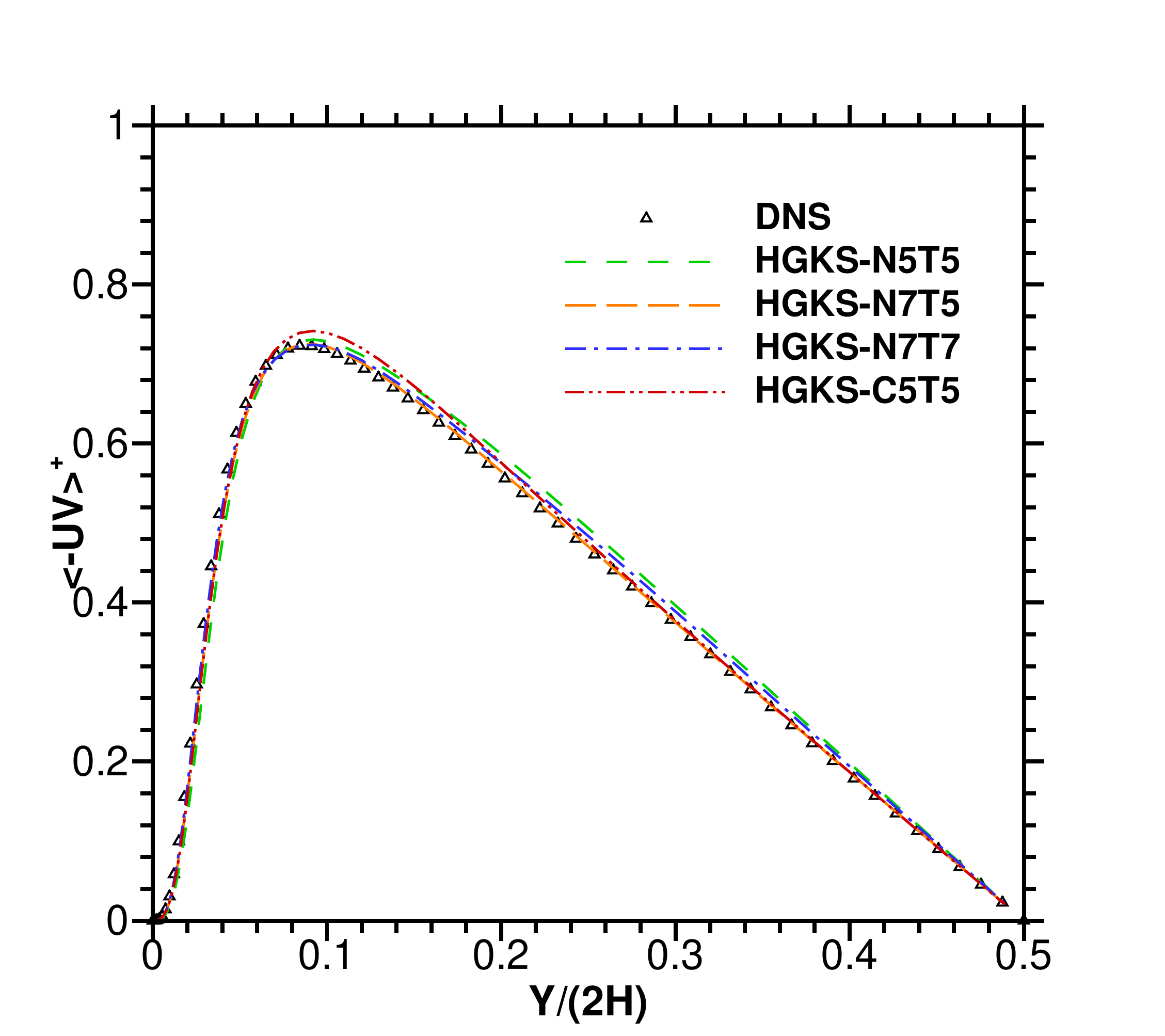}
    }   
     
\caption{\label{180_wrms}(a) Root-mean-square fluctuation velocity profiles in spanwise direction and (b) Reynolds stress profiles for $Re_\tau=180$.} 
	
\end{figure*}

For the channel flow at $Re_\tau=180$, the normalized root-mean-square fluctuation velocity profiles in the streamwise and normal-boundary directions are shown in Fig. \ref{180_urms}. In Fig. \ref{180_wrms}, the root-mean-square fluctuation velocity profiles for the spanwise direction and the normalized Reynolds stress profiles are presented.
In terms of the solutions of the root-mean-square fluctuation velocity in three directions and the Reynolds stress, the higher-order schemes improve the accuracy of the solution significantly both in near wall and near center-line region. Additionally, the results from the 5th-order compact scheme appear better than that from the 5th-order non-compact scheme.

\begin{figure*} [!h]
\centering
    \subfigure[]{
        \includegraphics[height=0.4\textwidth]{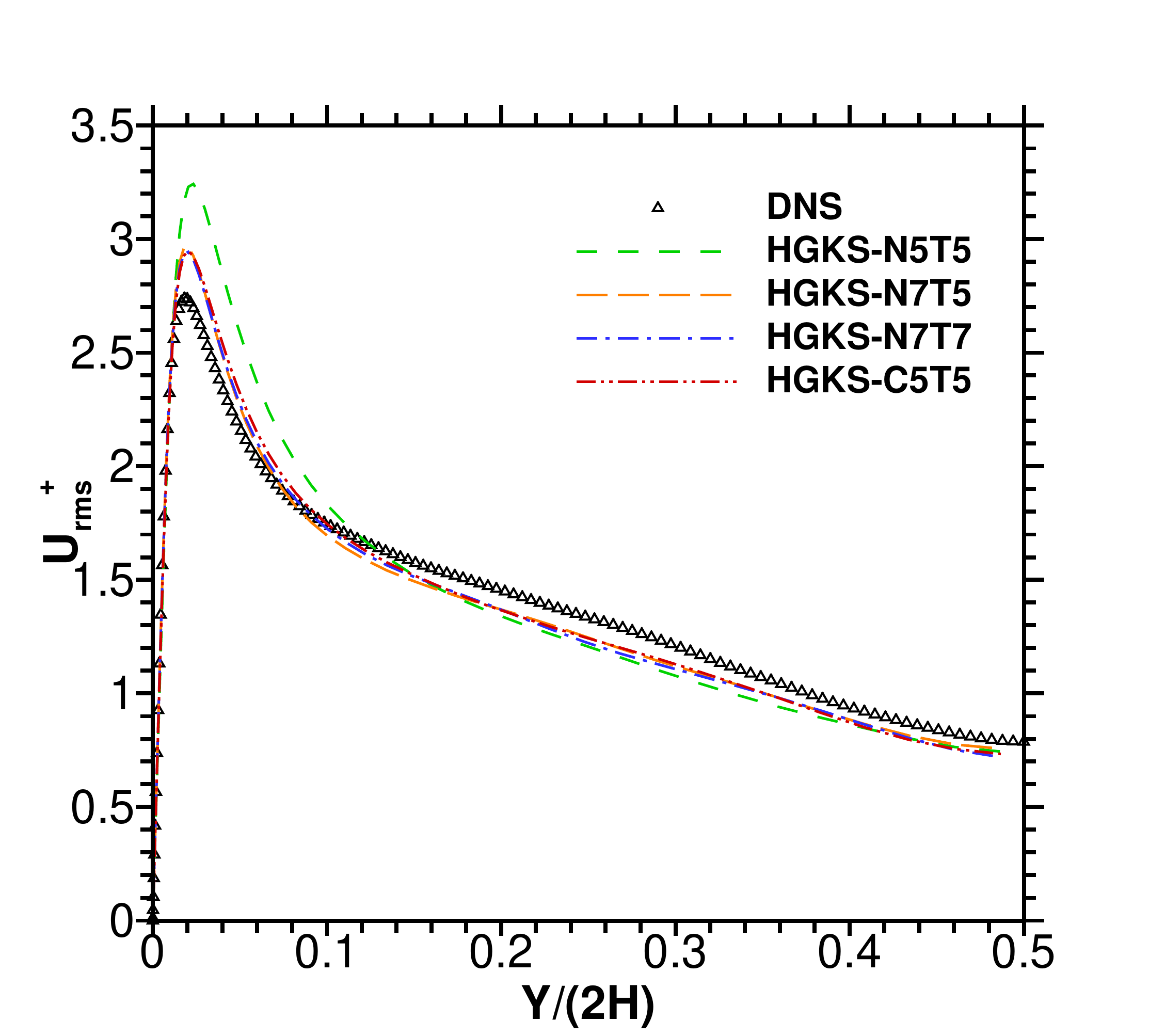}
    }
    \subfigure[]{
        \includegraphics[height=0.4\textwidth]{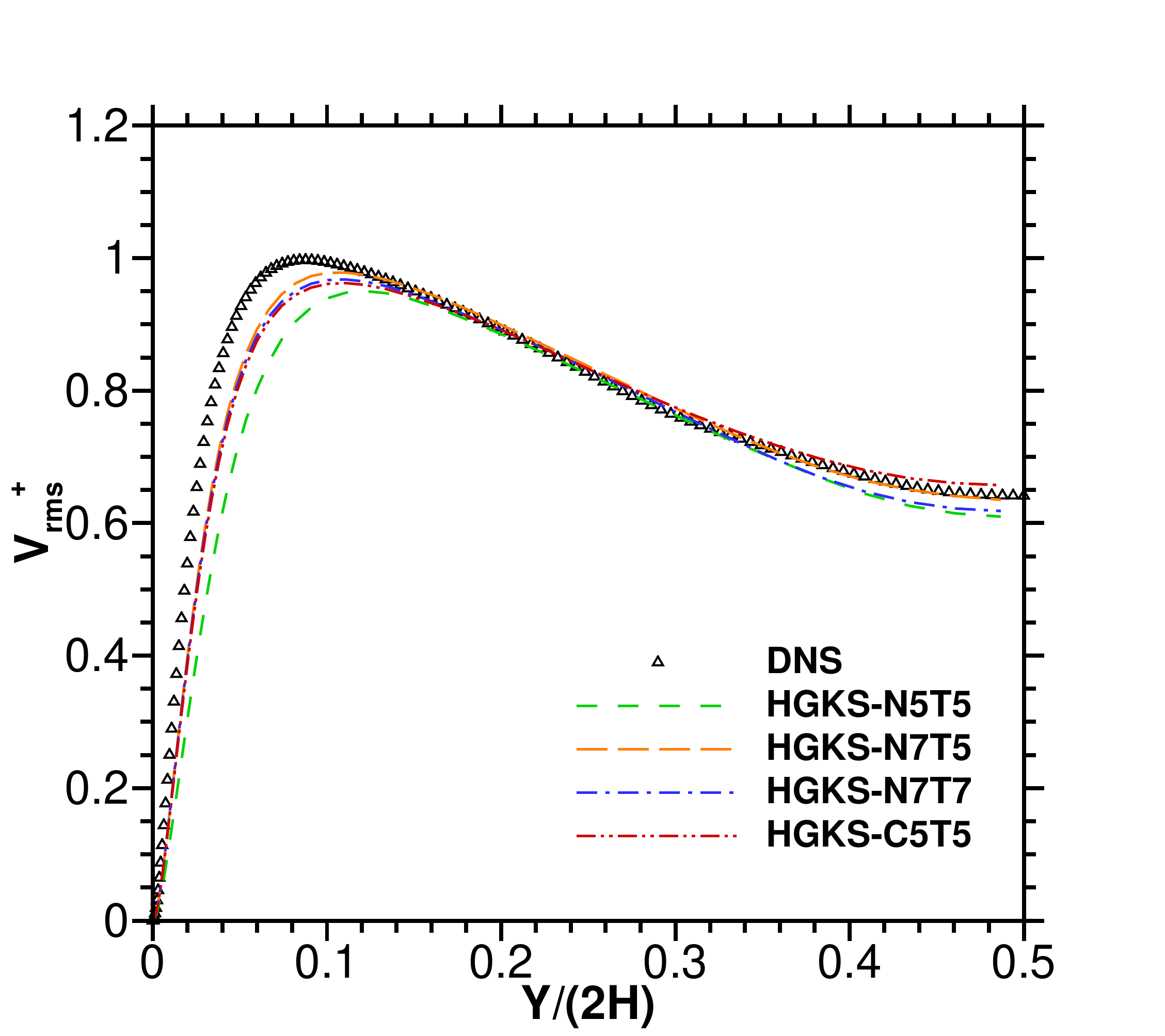}
    }   
	\caption{\label{395_urms} Root-mean-square fluctuation velocity profiles in (a) streamwise and (b) normal-boundary directions for $Re_\tau=395$.}	
%
%
    \subfigure[]{
        \includegraphics[height=0.4\textwidth]{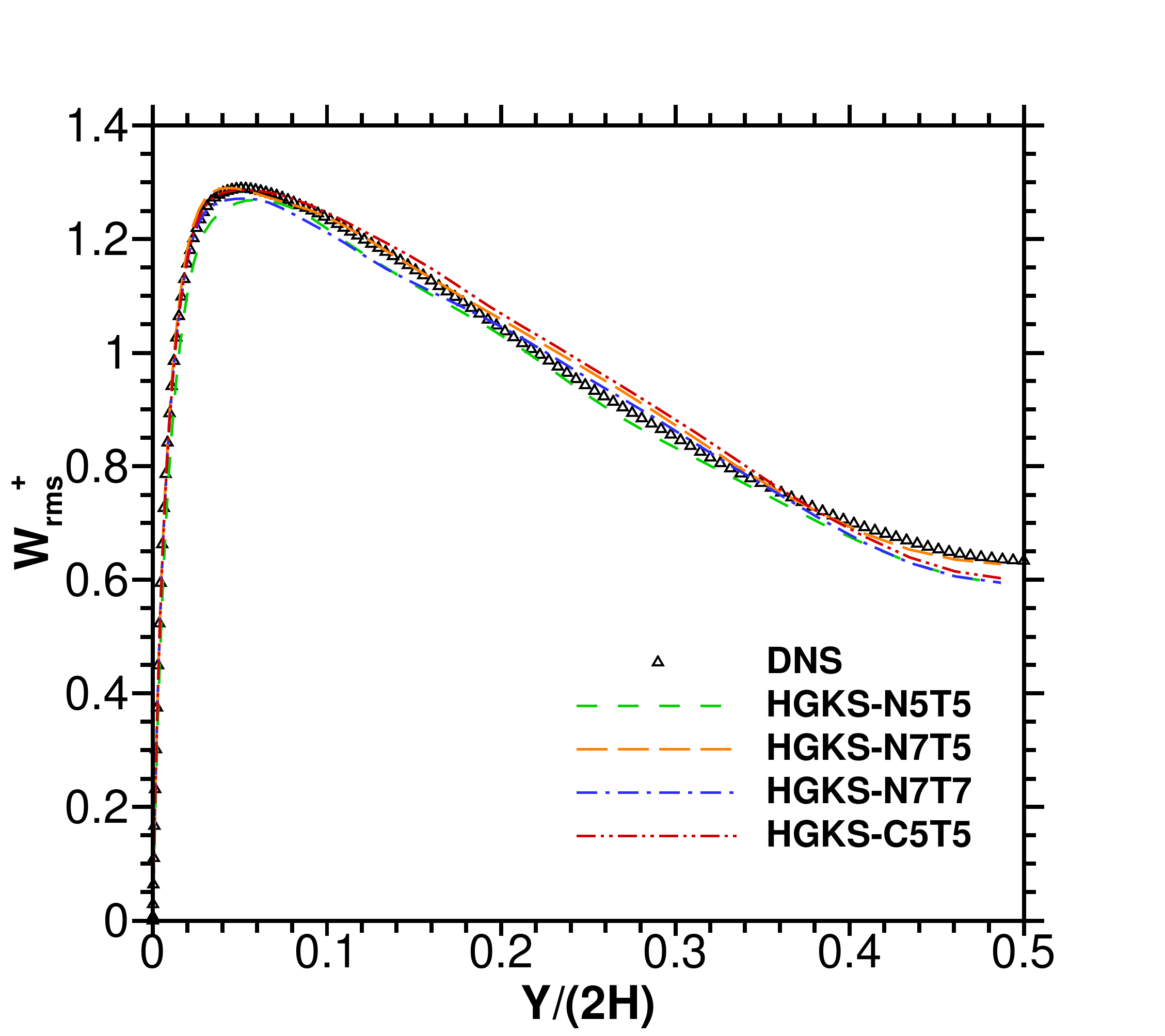}
    }
    \subfigure[]{
        \includegraphics[height=0.4\textwidth]{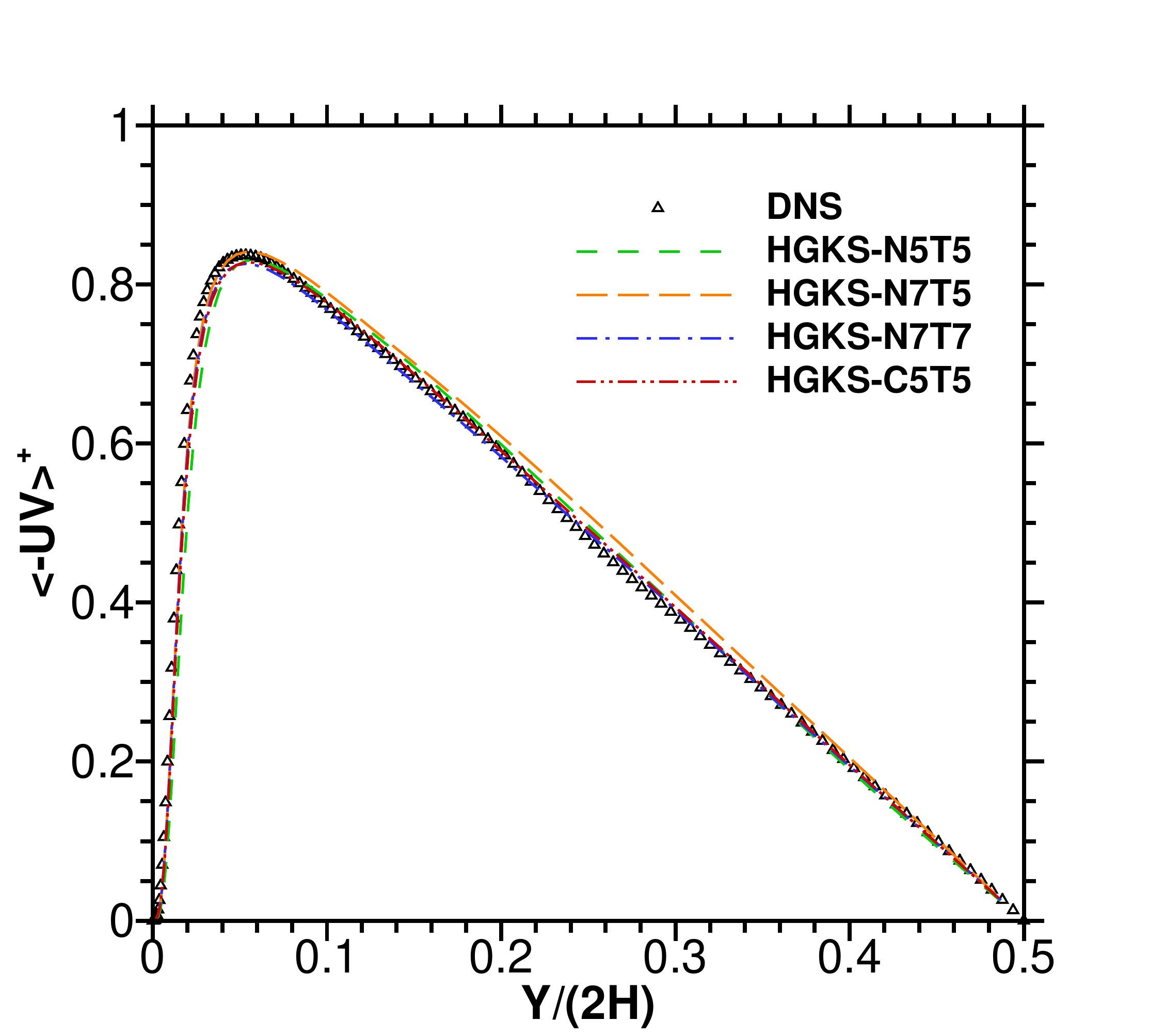}
    }   
     
\caption{\label{395_wrms}(a) Root-mean-square fluctuation velocity profiles in spanwise direction and (b) Reynolds stress profiles for $Re_\tau=395$.} 
	
\end{figure*}

Fig. \ref{395_urms} shows the normalized root-mean-square fluctuation velocity profiles in the streamwise and normal-boundary directions at $Re_\tau = 395$. It can be observed that the results of 7th-order non-compact scheme and 5th-order compact scheme are all closer to the results of DNS than 5th-order non-compact scheme. The 5th-order compact scheme can almost achieve the accuracy of the 7th-order non-compact scheme. 
The root-mean-square fluctuation velocity profiles for the spanwise direction and the normalized Reynolds stress profiles are shown in Fig. \ref{395_wrms}. For those results, all the schemes can match the DNS solution well.  



\subsubsection{Energy spectra}
\begin{figure*} [!h]
\centering
    \subfigure[streamwise-direction velocity]{
        \includegraphics[height=0.27\textwidth]{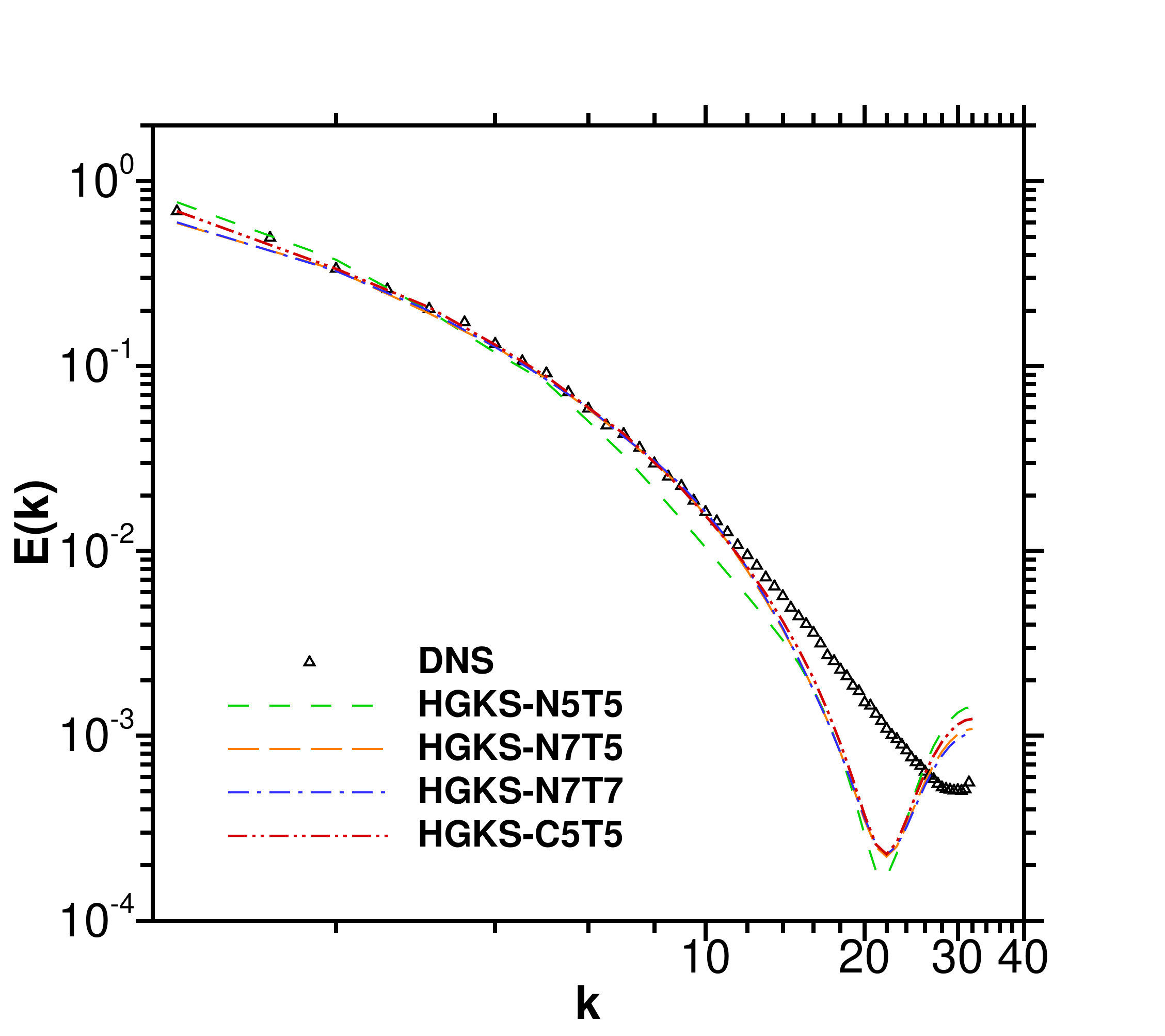}
    }
    \subfigure[normal-boundary-direction velocity]{
        \includegraphics[height=0.27\textwidth]{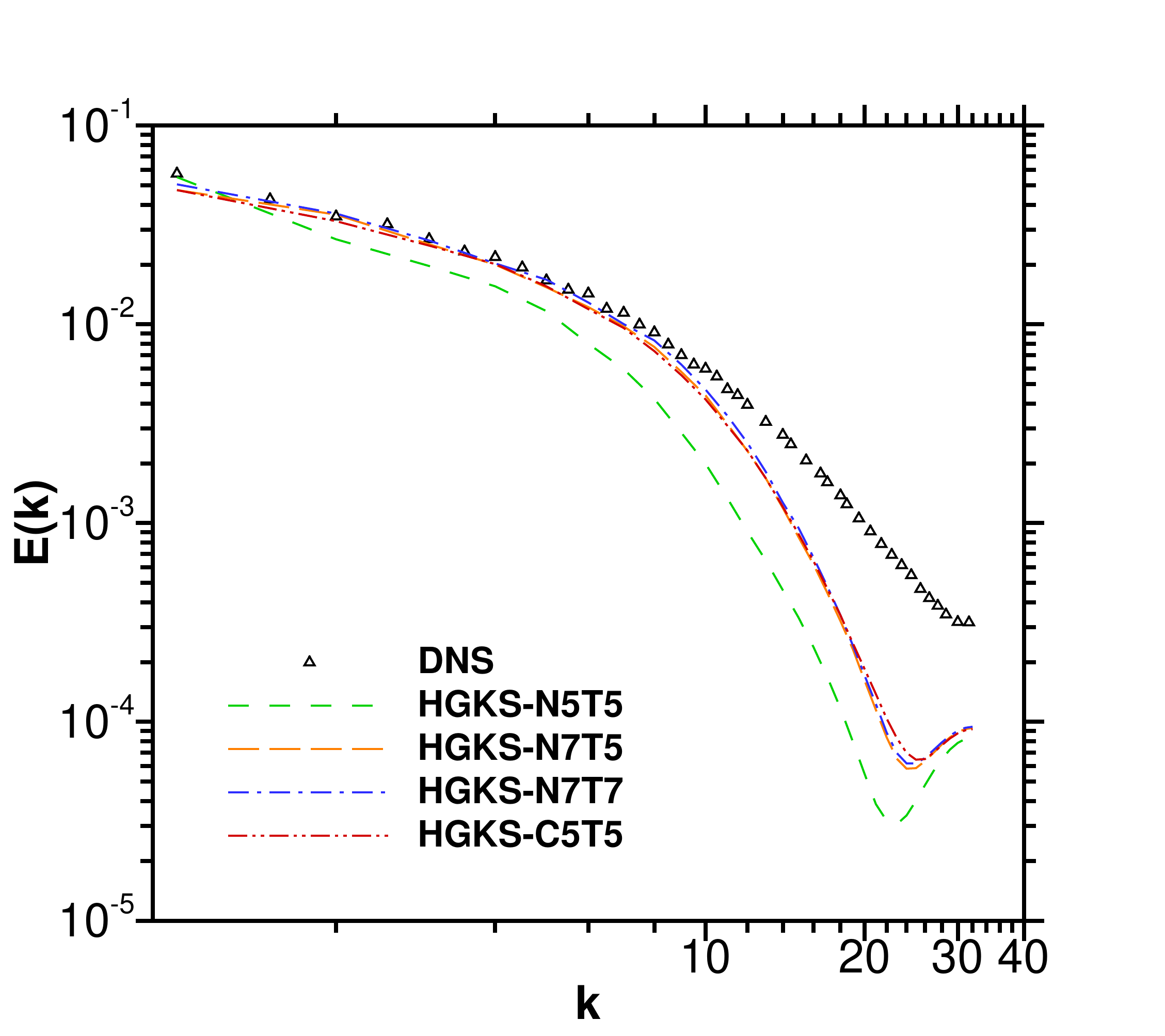}
    }   
       \subfigure[spanwise-direction velocity]{
        \includegraphics[height=0.27\textwidth]{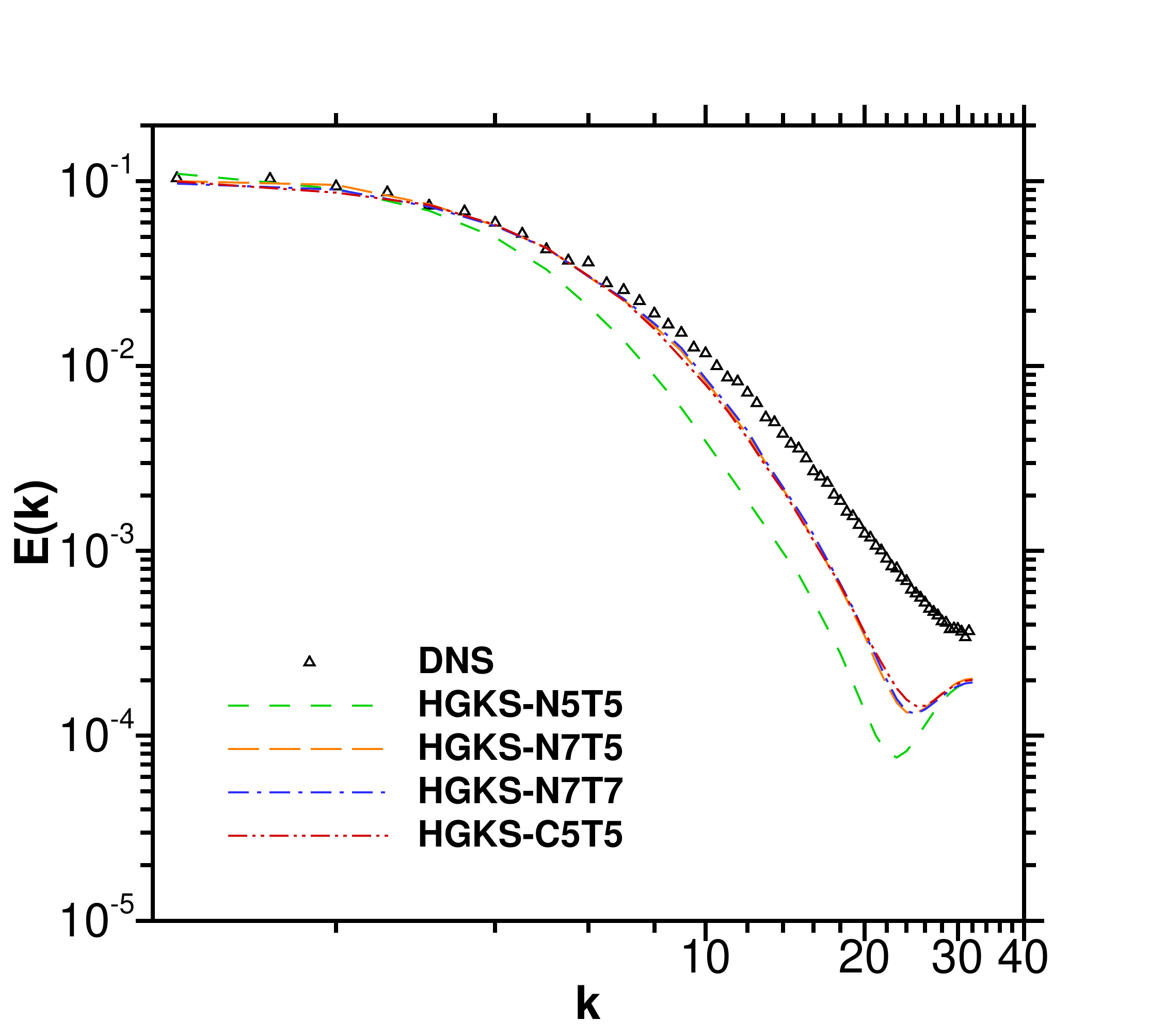}
    }

    \subfigure[streamwise-direction velocity]{
        \includegraphics[height=0.27\textwidth]{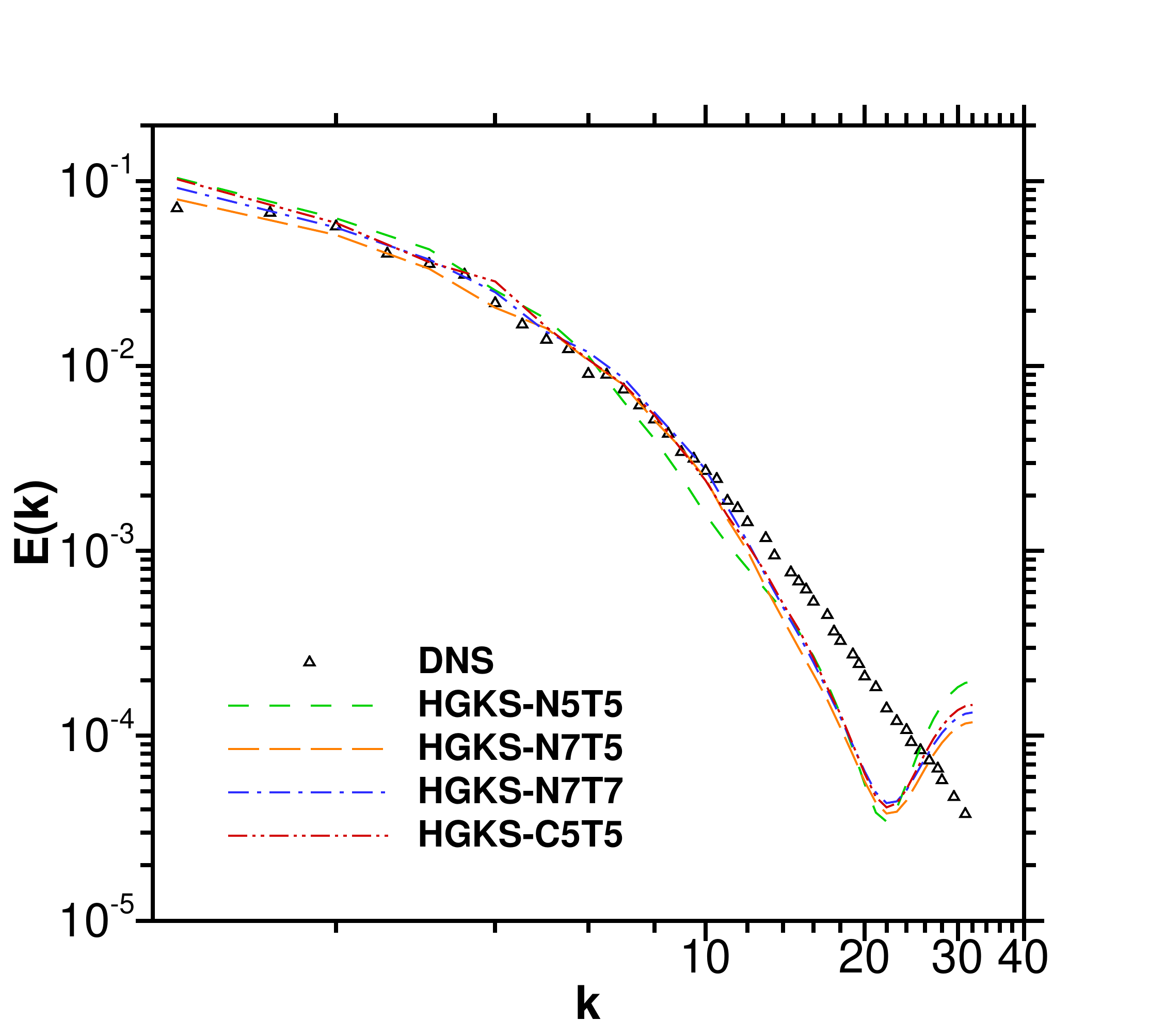}
    }
    \subfigure[normal-boundary-direction velocity]{
        \includegraphics[height=0.27\textwidth]{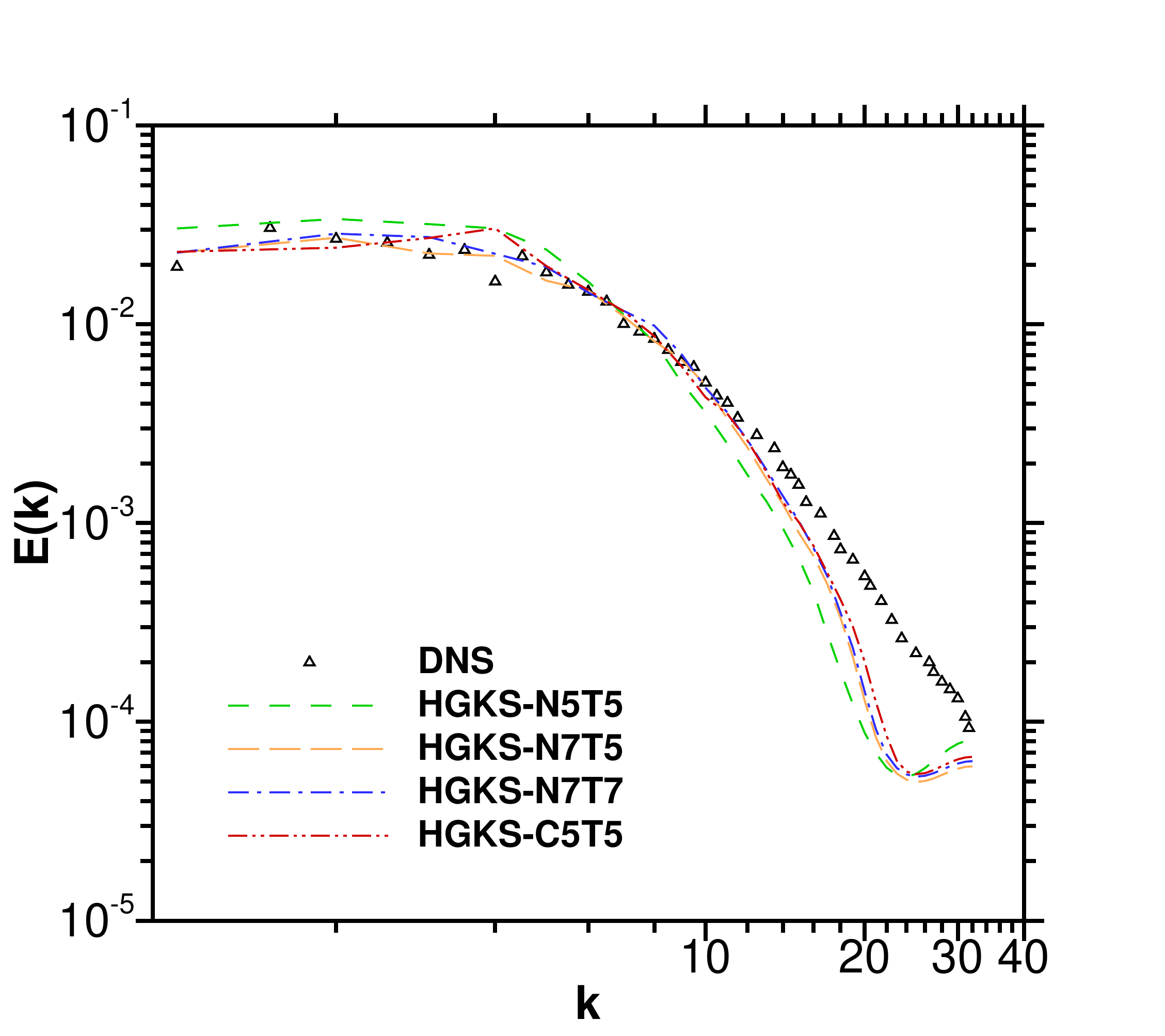}
    }   
       \subfigure[spanwise-direction velocity]{
        \includegraphics[height=0.27\textwidth]{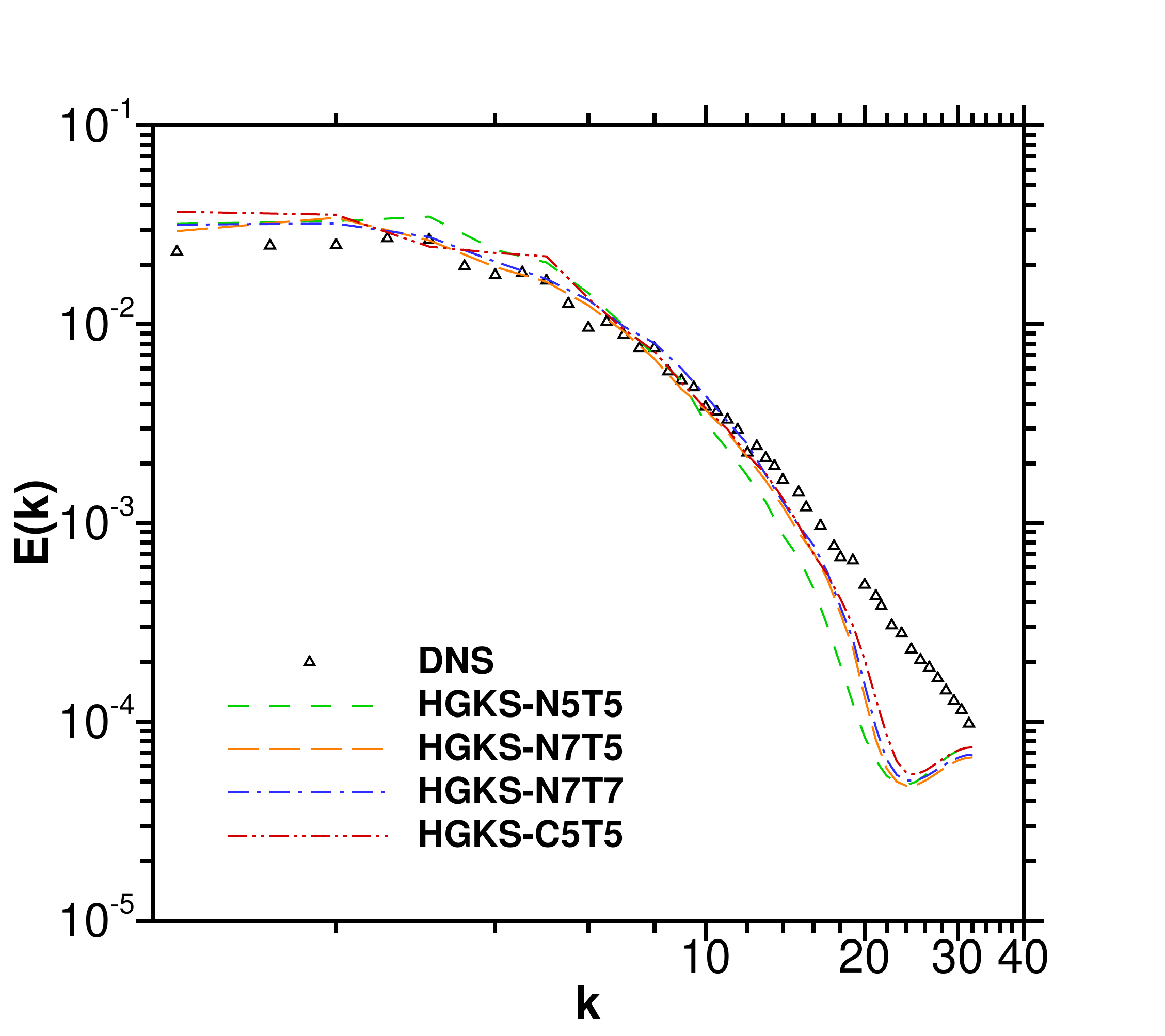}
    }   
	\caption{\label{180_ener_liu} Energy spectra at y+ = 30 (above) and y+ = 180 (below) in streamwise direction for $Re_\tau=180$.}	
\end{figure*}

\begin{figure*} [!h]
\centering
    \subfigure[streamwise-direction velocity]{
        \includegraphics[height=0.27\textwidth]{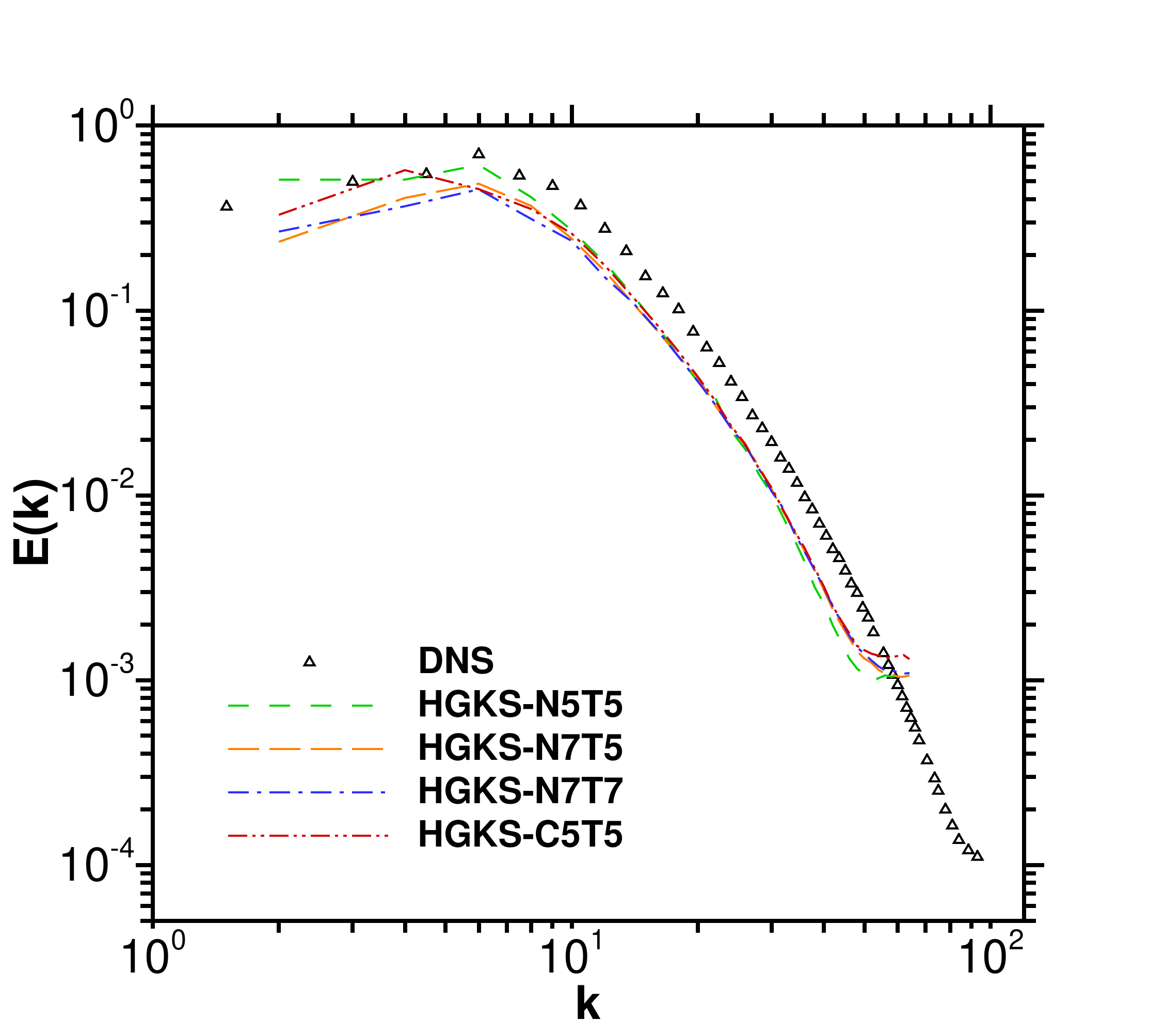}
    }
    \subfigure[normal-boundary-direction velocity]{
        \includegraphics[height=0.27\textwidth]{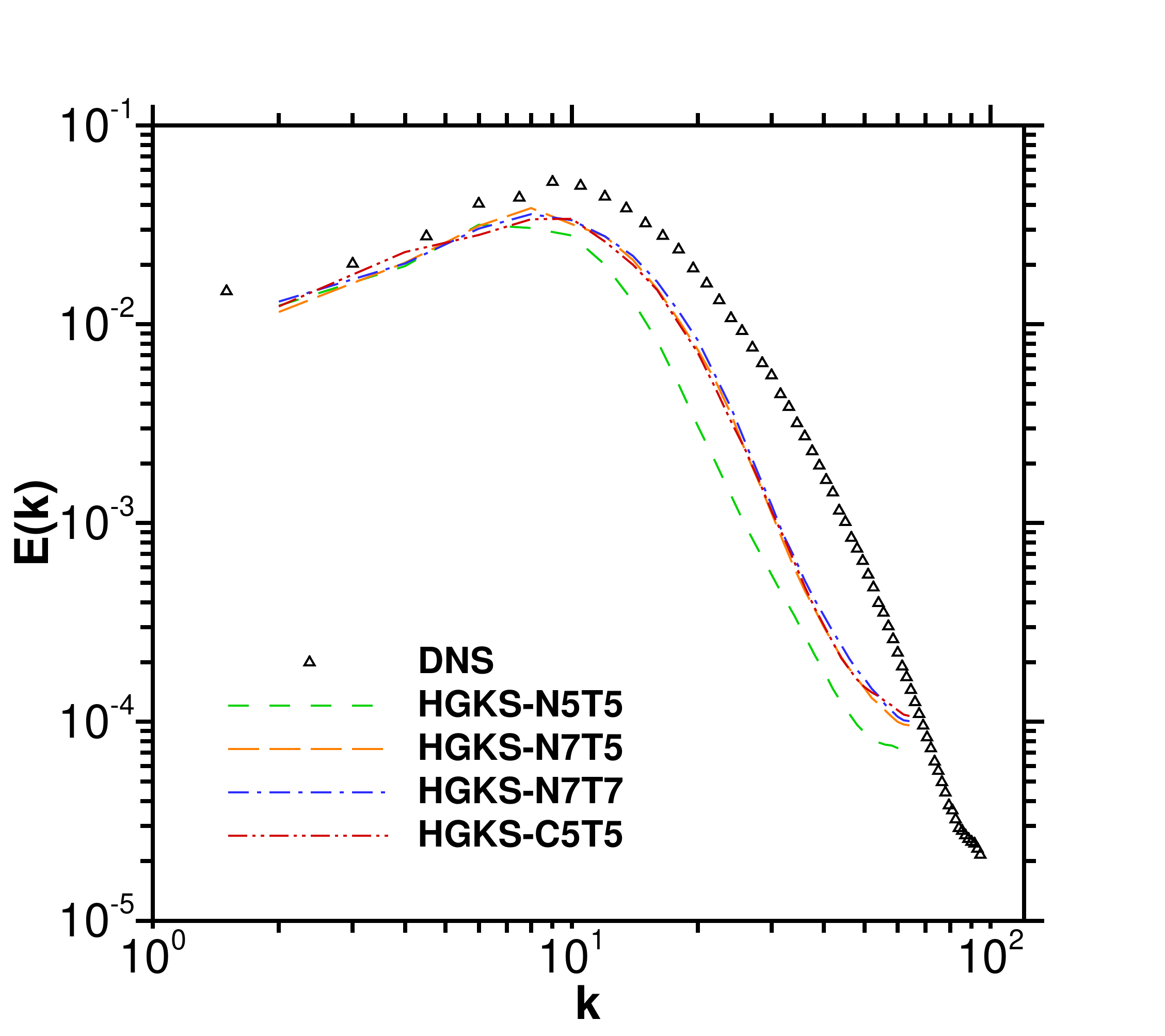}
    }   
       \subfigure[spanwise-direction velocity]{
        \includegraphics[height=0.27\textwidth]{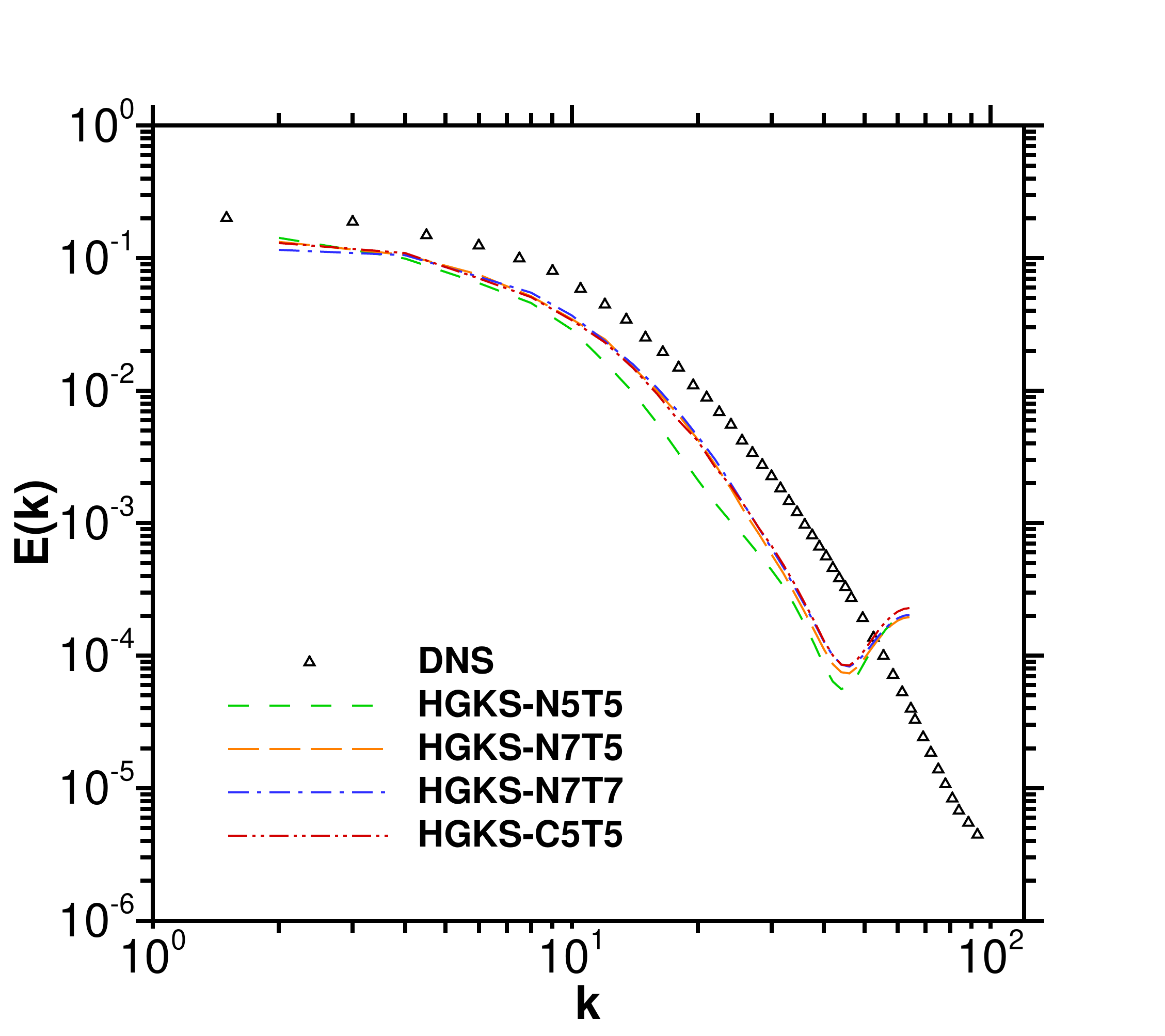}
    }

    \subfigure[streamwise-direction velocity]{
        \includegraphics[height=0.27\textwidth]{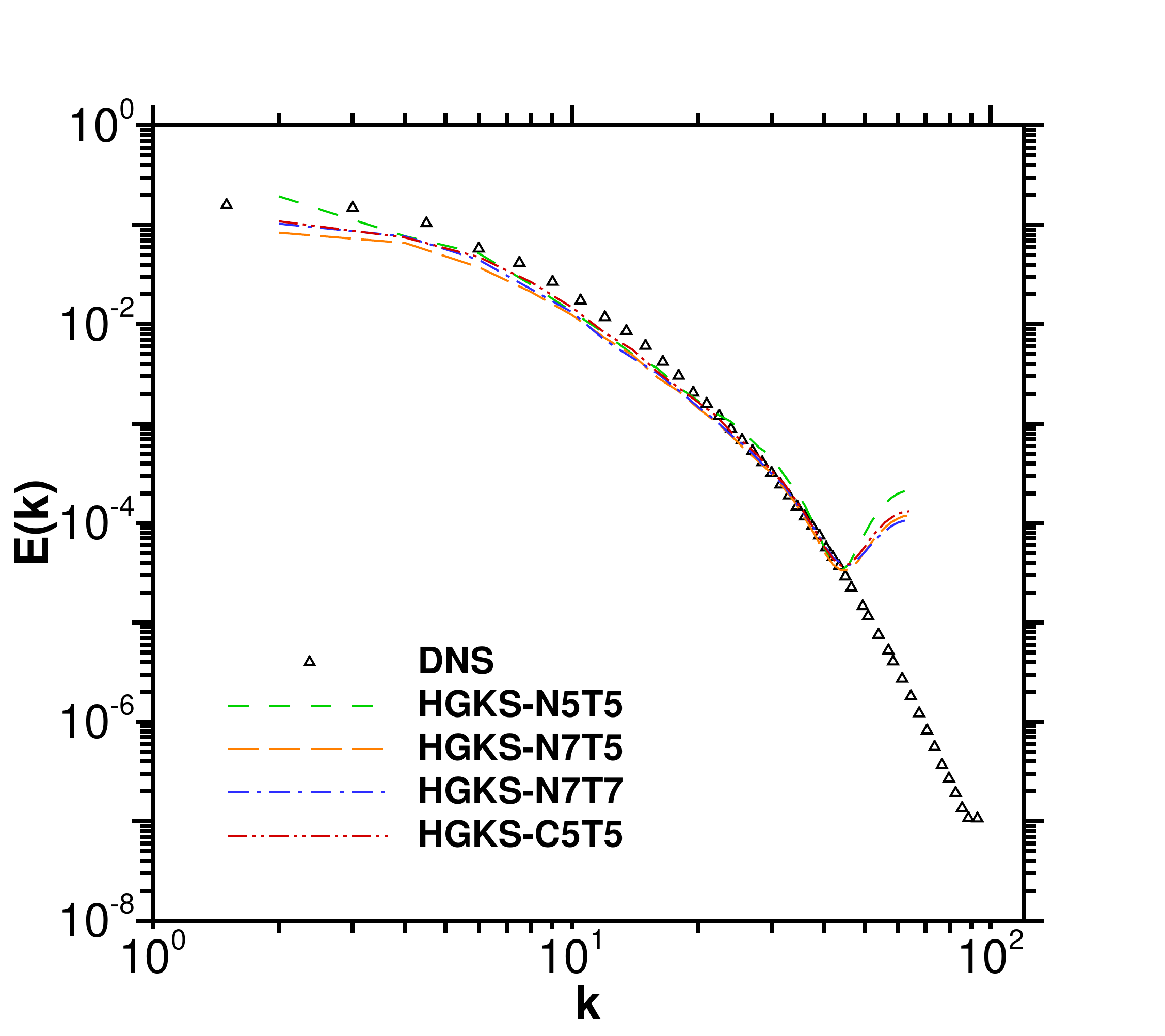}
    }
    \subfigure[normal-boundary-direction velocity]{
        \includegraphics[height=0.27\textwidth]{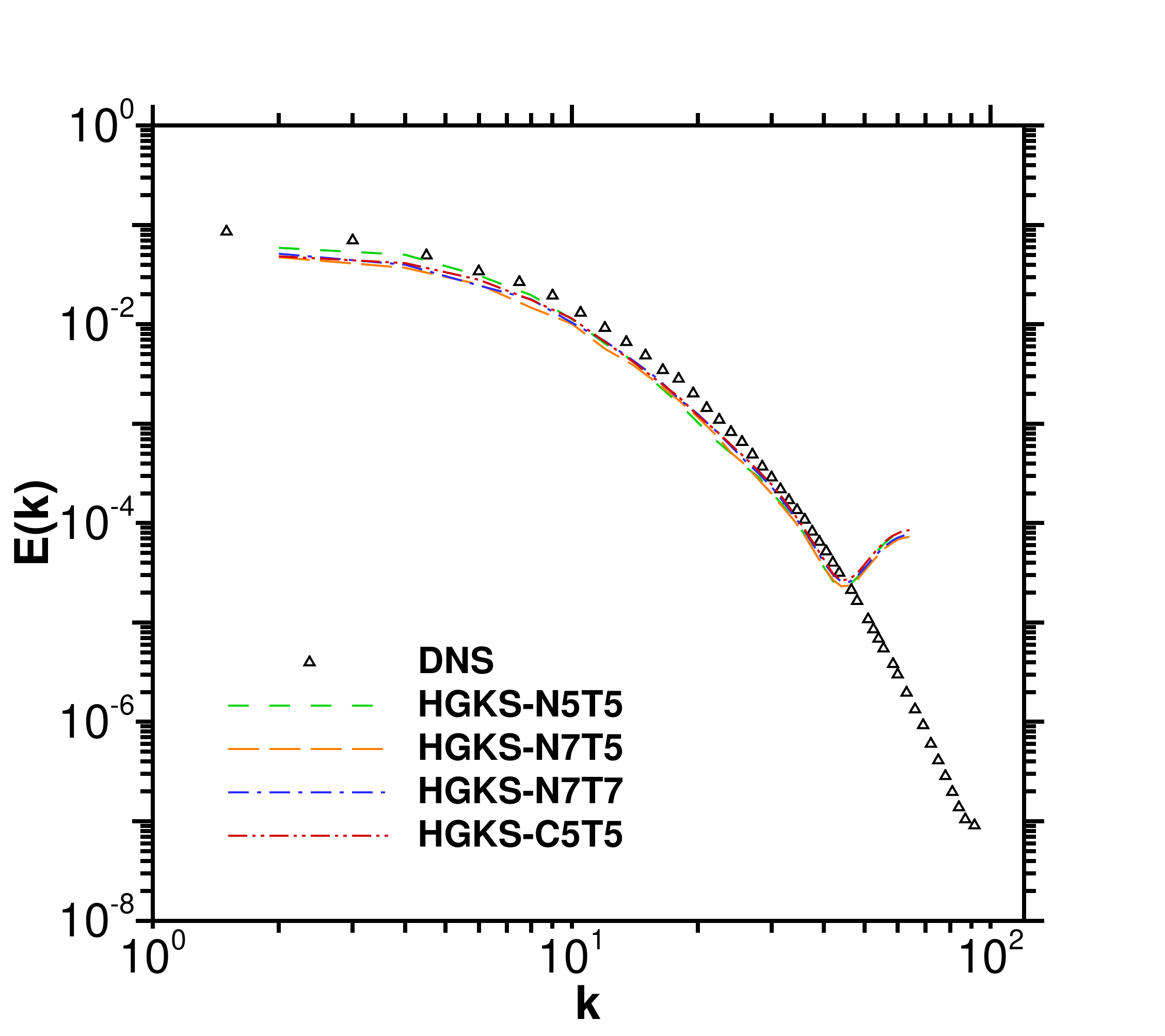}
    }   
       \subfigure[spanwise-direction velocity]{
        \includegraphics[height=0.27\textwidth]{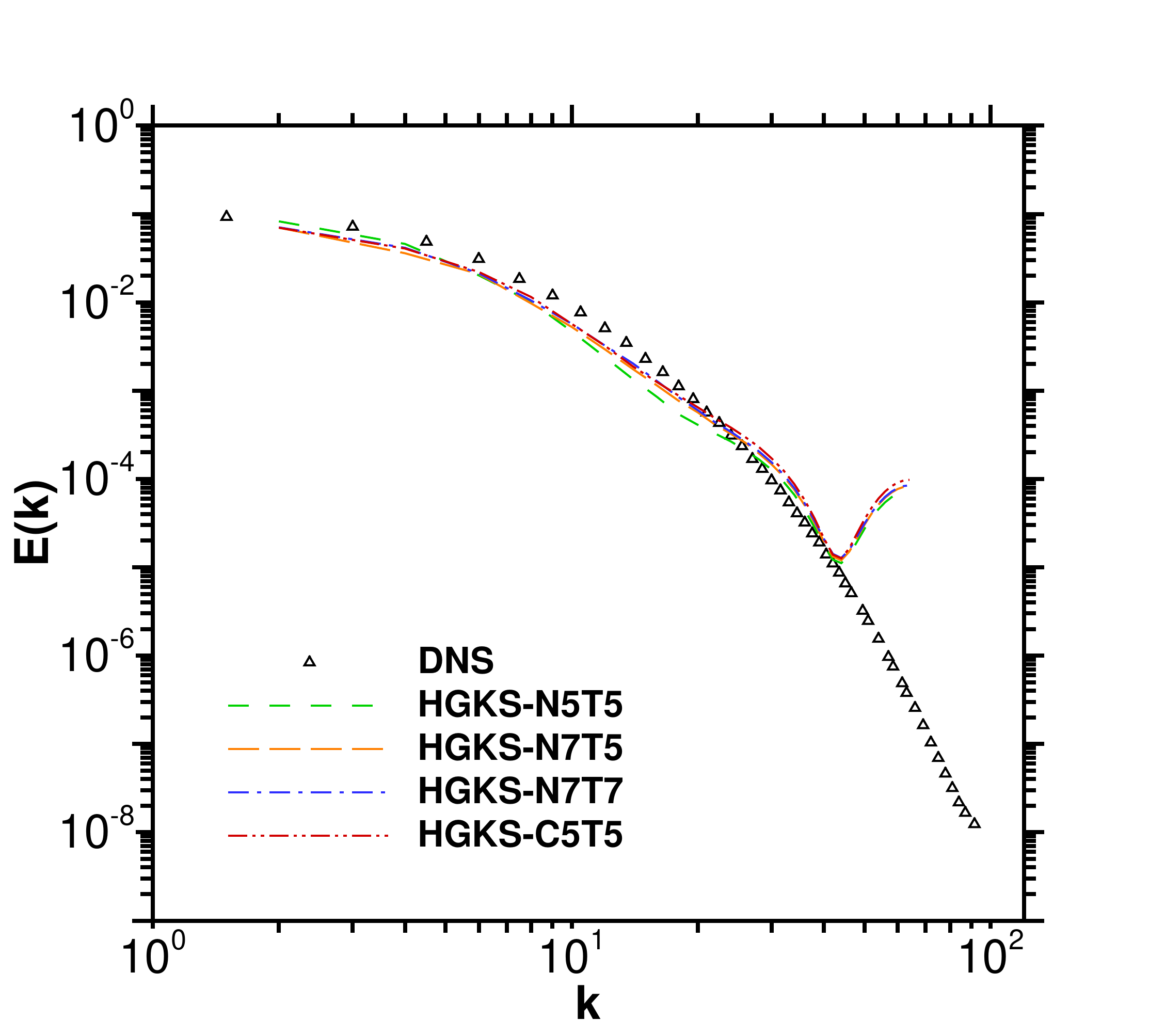}
    }   
	\caption{\label{180_ener_zan} Energy spectra at y+ = 30 (above) and y+ = 180  (below) in spanwise direction for $Re_\tau=180$.}	
\end{figure*}

Figs. \ref{180_ener_liu} and \ref{180_ener_zan}  show the energy spectra of the fluctuating velocity components in the streamwise and spanwise directions of the turbulent channel flow at $Re_\tau=$ 180 for $y^+=$ 30 and 180. In the low wavenumber region, all ILES results can match with DNS results well. 
We observe that the energy spectra at larger wavenumber is preserved by both 7th-order non-compact scheme and 5th-order compact scheme. Thus, the higher-order scheme and compact scheme can better resolve smaller scale turbulent structures.
 Additionally, the results in the near wall region ($y^+=30$) are improved more obviously than those in the near center region ($y^+=180$).
However, there is still a gap between ILES results and DNS results in the high wavenumber region. Furthermore, an unnatural leveling-off of the energy spectrum near the cutoff wavenumber is apparent for all the ILES results; this is because the mesh resolution in this area is too low.
The results of energy spectra of the fluctuating velocity component for the turbulent channel flow at $Re_\tau=$ 395 show a similar behavior and are omitted here. 

\subsubsection{Turbulence structure}

\begin{figure*} [!h]
\centering
    \subfigure[]{
        \includegraphics[height=0.4\textwidth]{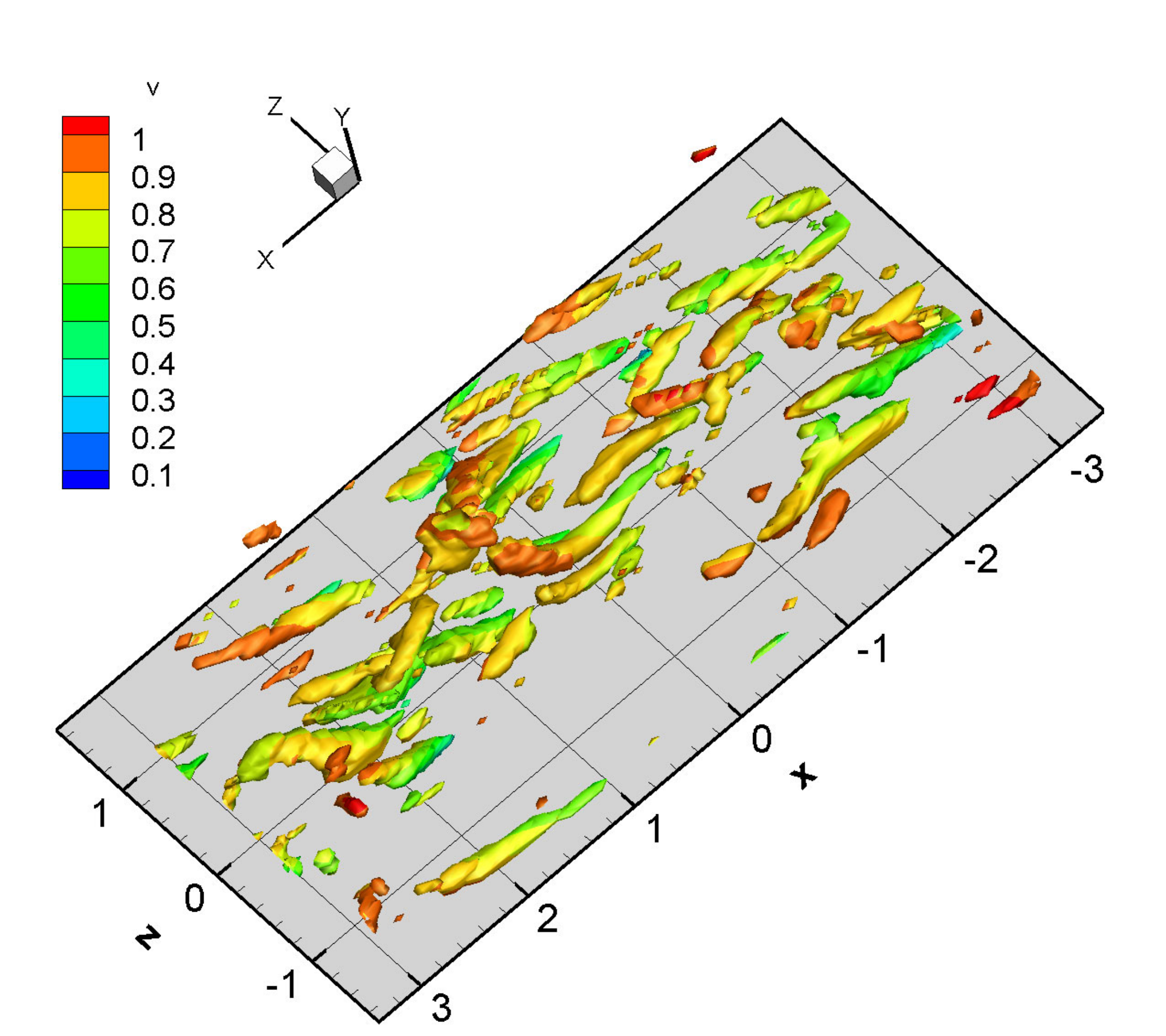}
    }
    \subfigure[]{
        \includegraphics[height=0.4\textwidth]{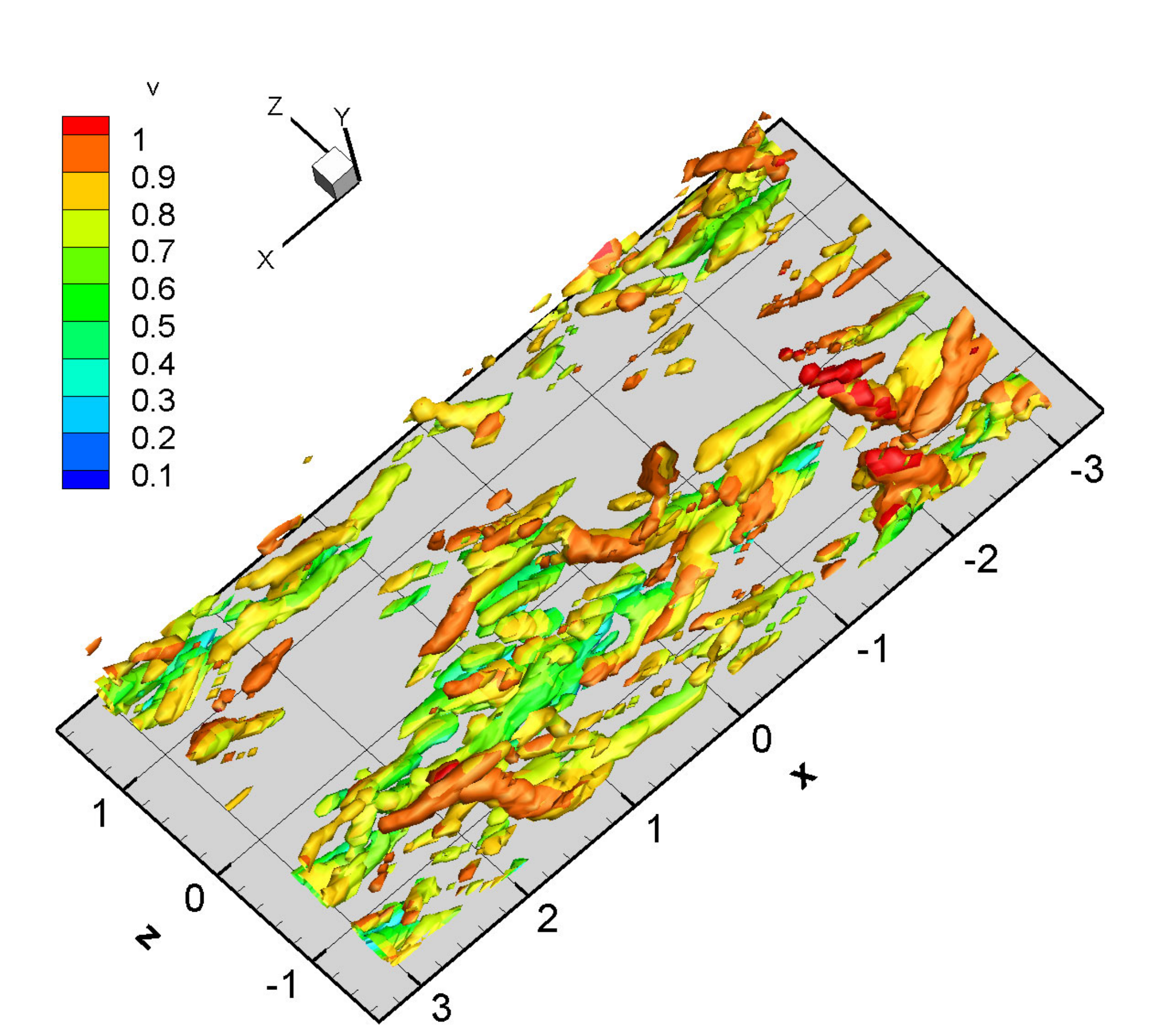}
    }   
	\caption{\label{180_n5} Q-criterion iso-surfaces for (a) HGKS-N5T5 and (b) HGKS-N7T5 (iso-value = 0.5 colored by streamwise velocity) for $Re_{\tau}=180$ channel flow.}	

    \subfigure[]{
        \includegraphics[height=0.4\textwidth]{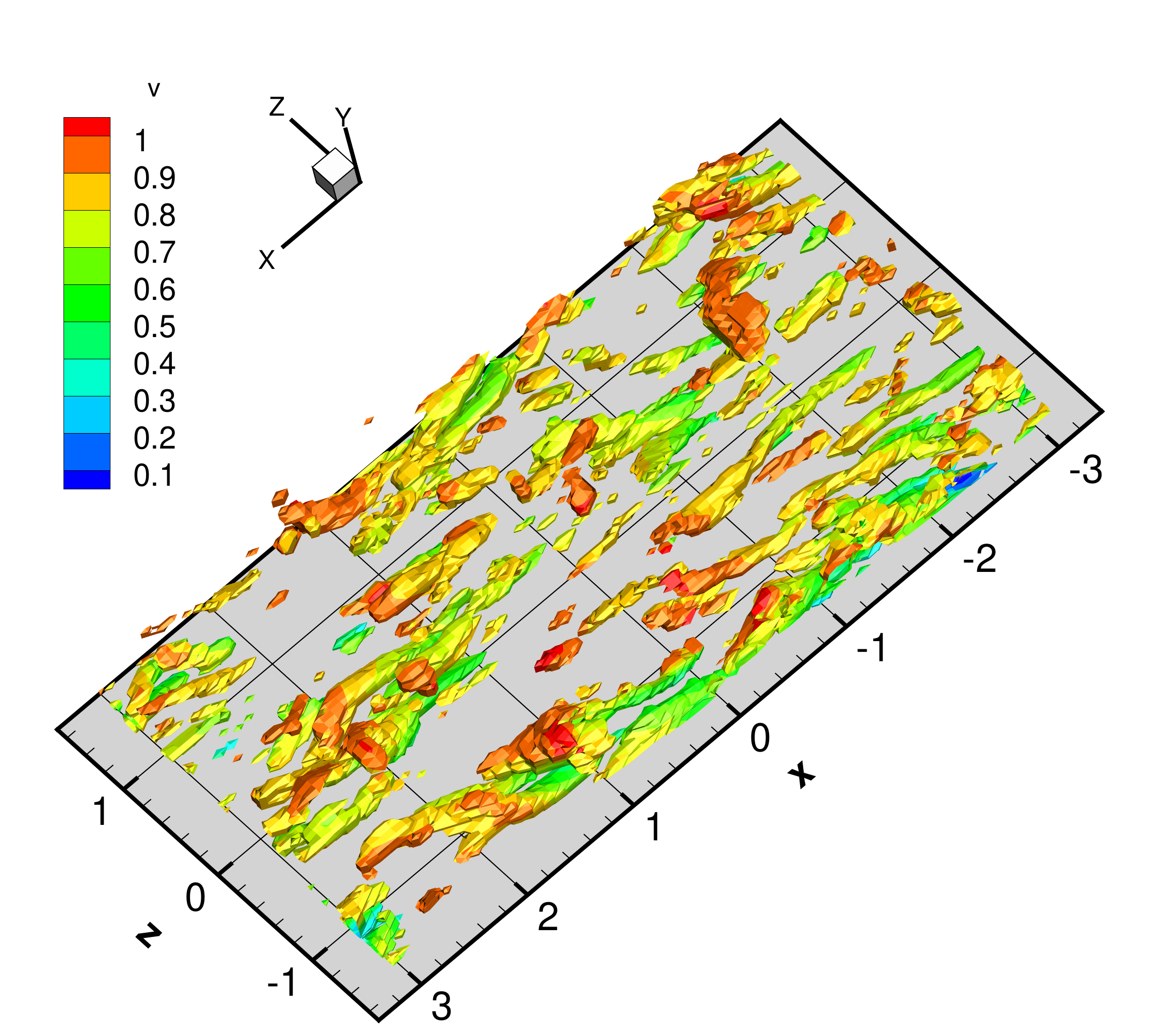}
    }
    \subfigure[]{
        \includegraphics[height=0.4\textwidth]{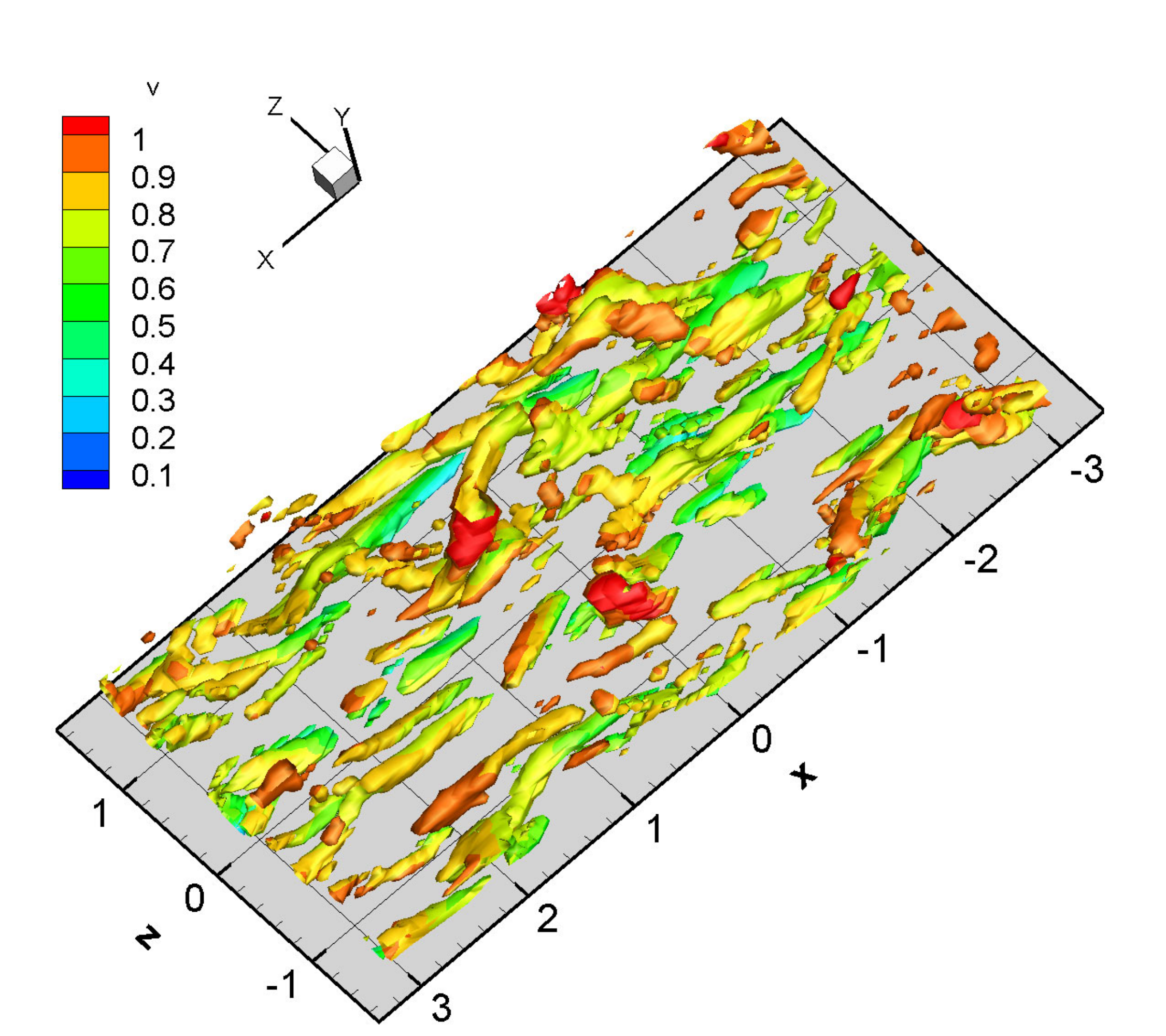}
    }   
     
\caption{\label{180_n7} Q-criterion iso-surfaces for (a) HGKS-N7T7 and (b) HGKS-C5T5 (iso-value = 0.5 colored by streamwise velocity) for $Re_{\tau}=180$ channel flow.} 
	
\end{figure*}

Figs. \ref{180_n5} and \ref{180_n7} show the Q-criterion iso-surfaces of the
turbulent channel flow at $Re_{\tau}=180$. As noted above, the Q-criterion
iso-surface indicates the ability of a numerical scheme to resolve
turbulent structures. A higher-accuracy scheme corresponds to the
resolution of more turbulent structures. We choose the approximate first peak point of Reynolds number after 400 characteristic periodic time ($T=H/U_c$) as the comparison time point. Specifically, the selected time points for HGKS-N5T5, HGKS-N7T5, HGKS-N7T7, and HGKS-C5T5 are 414 $T$, 415 $T$, 400 $T$, and 410 $T$ respectively.
It is shown that schemes with the higher-order non-compact reconstruction and the compact reconstruction resolve more vortex structures than 5th-order non-compact HGKS.
It can also be observed that vortices located above the low-speed streaks are ejected away from the wall and elongated in streamwise, which produces hairpin vortices stretched by the ambient shear. 
This phenomenon have also been reported by previous ILES studies \cite{2002Large}. 
The results of Q-criterion iso-surface for the turbulent channel flow at $Re_\tau=$ 395 shows a similar behavior and are omitted here. 

%
%

\section{\label{sec6} Discussion and conclusion}
In this study, we develop the HGKS with 7th-order non-compact reconstruction in the normal and tangential directions, as well as the HGKS with 5th-order compact reconstruction in the normal direction, which are used to investigate the performance of these schemes in turbulence simulation. We apply the direction by direction reconstruction strategy on rectangular mesh in the three-dimensional simulation to get the values of macroscopic conserved quantities, and use the analytical solution of the kinetic model equation to get flux function at the Gaussian points on the cell interfaces.

 Firstly, we test the three-dimensional advection of density perturbation to verify the accuracy of the code. With 16 Gaussian points on the cell-interface, the 7th-order accuracy can be obtained, but when 4 Gaussian points are used, the order cannot be maintained, and the accuracy will fall to 4th-order with mesh refinement. This illustrates that for high-order schemes, sufficient Gaussian points are needed to maintain high-order accuracy.

Then we apply the high-order scheme and the compact scheme for turbulence simulation, in which three-dimensional TGV and turbulent channel flows at $Re_\tau$=180 and 395 are tested. If 16 Gaussian points are used for turbulence simulation, the computational cost is too high. So we still use 4 Gaussian points in the simulation. Various results are obtained for the qualitative and quantitative comparisons. The results obtained from 7th-order non-compact and  5th-order compact schemes are much more accurate than that from the 5th-order non-compact scheme.
 For the TGV case, the time history of kinetic energy, dissipation rate and enstrophy are compared.
The results show that that both the high-order reconstruction and compact reconstruction can improve the numerical results. Increasing the reconstruction order in the tangential direction can also improve the results.
 The time history of numerical dissipation from different schemes is investigated. When increasing the order of the reconstruction or using compact reconstruction, the numerical dissipation decreases. The 5th-order scheme for ILES is still over-dissipative.
For turbulent channel flows, the mean velocity profiles, Reynolds stress, energy spectra and Q-criterion iso-surfaces are compared among different schemes.
The results of 5th-order compact scheme are close to those of the 7th-order non-compact scheme.
Especially, compared with the DNS solution, the energy spectra at larger wavenumber is preserved by both 7th-order non-compact scheme and 5th-order compact scheme. In general, the compact scheme can resolve smaller-scale turbulent structures better.

Overall, we develop the HGKS with high-order non-compact and compact reconstruction in three-dimension and apply the schemes in ILES. 
For the high-order scheme, the numerical dissipation is reduced, and the resolution is increased.
HGKS with compact reconstruction has smaller stencils. The compact reconstruction has the consistent physical and numerical domains of dependence. The compact GKS has a favorable performance for turbulence simulation in resolving its multi-scale structure.

In the future work, we will investigate the performance of high-order non-compact  and compact reconstruction of HGKS for ILES at high Reynolds numbers and high Mach numbers.\\

\textbf{Declaration of Competing Interest}\\

The authors declare that they have no known competing financial interests or personal relationships that could have appeared to influence the work reported in this paper.\\

\textbf{Acknowledgments}\\

This work was supported by the National Natural Science Foundation of China (NSFC; Grant Nos. 12172316 and 12172161) and Hong Kong research grant council (16208021 and 16301222).

\bibliographystyle{unsrt}
\bibliography{aipsamp}
\end{document}